\documentclass[twocolumn]{aastex631}  
\usepackage{enumitem}
\usepackage{xcolor}
\usepackage{amsmath} 
\usepackage{xcolor, soul}
\sethlcolor{yellow}
\usepackage{subfigure}
\usepackage{tablefootnote}
\usepackage{bm}
\hypersetup{backref=true,pagebackref=true,  hyperindex=true,colorlinks=blue,breaklinks=true,  urlcolor=violet,linkcolor= red, bookmarks=true, 	bookmarksopen=true,  filecolor=cyan, citecolor=blue, linkbordercolor=blue}
\shorttitle{Probing periodic trends in the TESS light curves of known DWDs}
\shortauthors{Sajadian,  Asadi}

\begin{document}
\title{Probing periodic trends in the TESS light curves of the seventeen known Double White Dwarf systems}

\author[0000-0002-0167-3595]{Sedighe Sajadian}
\affiliation{Department of Physics, Isfahan University of Technology, Isfahan 84156-83111, Iran, \url{s.sajadian@iut.ac.ir}}

\author{Aref Asadi}
\affiliation{Department of Physics, Isfahan University of Technology, Isfahan 84156-83111, Iran}

\begin{abstract}
There is a relatively large population of known double white dwarfs (DWDs) that were mostly discovered through spectroscopic observations and by measuring their radial velocity variations. Photometric observations from these systems give us additional information about their faint components by manifesting eclipsing or lensing signals or periodic trends such as ellipsoidal variations or Doppler boosting. To find these signals and trends we probe the public photometric data collected by the Transiting Exoplanet Survey Satellite (TESS) telescope from 17 known DWD systems. We use the Singular Spectrum Analysis (SSA) technique to de-noise their light curves. For DWD systems J1717$+$6757, J1557$+$2823, LP400$-$22, J1449$+$1717, J2132$+$0754, and J2151$+$1614 we find regular and periodic trends in their TESS light curves. The periodic trend in light curve J1449$+$1717 is caused by the blending effect due to a variable and bright star close to it which are unresolvable in the TESS observations. The discovered periodic trend for J1717$+$6757 was recovered by the TESS data. The periodic trends in light curves of J1557$+$2823 and J2151$+$1614 have the False Alarm Probability (FAP) values $\simeq 14.8,~36.5\%$. So their detected trends are likely noises with non-orbital origins. Periods of trends in light curves of LP400$-$22 and J2132$+$0754 are the same as and half of orbital periods, respectively. We evaluate possible ranges for Doppler boosting and ellipsoidal variations's amplitudes for these targets. This study highlights the importance of TESS data for identifying periodic trends such as ellipsoidal or intrinsic variations rather than short eclipsing/lensing signals in DWD light curves specially bright targets with ignorable blending.
\end{abstract}

\section{Introduction}\label{sec1}
One type of binary systems in our galaxy is double white dwarfs (DWDs) which are composed of two white dwarfs. DWD systems are noteworthy objects, because they potentially are progenitors of Type Ia supernovae and can radiate gravitational waves at their merging times. Considering the fact that $97\%$ of stars finally evolve into WDs, DWD systems should be more common than double neutron stars or double black holes. In our galaxy, around $100$-$300$ million DWDs exist and they are generally classified into two groups DA and DB \citep{2001Nelemans,Marsh_2011,2017AAToonen}. Up to now several surveys including the Sloan Digital Sky Survey (SDSS, \citealp{2000AJSDSS}), the Gaia space telescope \citep{2016AAGaia}, the Large Sky Area Multi-Object Fiber Spectroscopic Telescope (LAMOST, \citealt{2006LAMOST}), have identified large samples of WDs \citep[see, e.g.,][]{SDSSWDcatalog2013,WDGaiacatalo2018,2019SDSSWDcat,2022MNRASGuo}. Among these reported WDs several DWD systems were identified through discerning two overlapping spectra in spectroscopic observations or by measuring radial velocity (RV) variations \citep[e.g., ][]{2024AADWDsSDSS,2015ApJGianninas,Rebassa2021GaiaDWD,2020kilicmnras,2024MNRASmunday}. Additionally, the Extremely Low Mass (ELM) survey \citep{Brown_2010_ELMSurvey} has focused on ELM WDs (WDs with masses less than $0.3 M_{\odot}$ in binary systems) and measures RV variations by doing time-series spectroscopy for these objects and could discover a large sample of DWDs \citep[see, e.g., ][]{2020ApJBrown}. 

Either time-series spectroscopic or dense photometric observations from these systems could manifest their binarity as well as cause measuring their physical parameters. In the regard of deriving the physical parameters of two WDs in DWD systems, the best-fitted models for their spectra will offer the effective temperature $T_{\rm{eff}}$, and the surface gravity $\log_{10}[g]$. In 2D space $T_{\rm{eff}}-\log_{10}[g]$ WDs fall on specified tracks depending on their masses, which makes extracting WDs' mass. The RV variations also give the orbital period $T$. However, in some DWD systems one component is very faint and has ignorable footprints on measured spectra. Hence, measuring its mass from modeling spectra is not always possible. For such systems, spectroscopic observations and RV measurements will offer a low limit on the mass of faint WDs. 

Indeed, in an isolated DWD system two components are rotating around their common center of mass in an orbital plane at the inclination angle $i$ with respect to the sky plane. The resulting light curve of these systems, as received by the observer, may contain variations due to eclipsing, self-lensing, finite-lens size, Doppler boosting, and ellipsoidal variations which all are periodic and have different durations and periods. These photometric variations mostly depend on the inclination angle of the orbital plane $i$, the orbital period, and the mass and compactness of two components. 

\noindent For instance, massive WDs in DWD systems will act as a gravitational lens when two components have similar lines of sight towards the observer, the so-called self-lensing, which causes periodic enhancements in stellar light curves \citep{1973AAMaeder,1997ChPhLQin}. During a self-lensing event, some part of the images' disk may be obscured by the lens object (referred to as finite-lens size or eclipsing) which alters the lensing profile over time \citep{2002ApJAgol,2016ApJHan,2024sajadianfinite}. Hence, for edge-on DWD systems measuring masses of faint components is possible by discerning periodic eclipsing or lensing signals in their light curves \citep[e.g.,][]{2021GrazaWilson,2023MNRASMunday,2024arXivJin,2025AJsajadian}.
    
\noindent Ellipsoidal variations are considerable mostly for close binary systems in which one component is distorted by the strong gravitational force of its companion, into a teardrop shape. This effect was realized in different binary systems including DWD ones \citep{1993ApJEllipsoidal}. Measuring this effect also helps to characterize the binary system \citep{2010Nicholls,2024arXiv240811100P,2022ApJSorebella}. Doppler boosting variations occur in edge-on and close binary systems \citep{2010DopplerBoosting,2015NaturDopplerB}. To discern any potential periodic trends or signals for known DWD systems, we extract their light curves from the public photometric data taken by the NASA's Transiting Exoplanet Survey Satellite\footnote{\url{https://exoplanets.nasa.gov/tess/}} (TESS, \citealt{Ricker2024}) telescope.

TESS is a space telescope designed to densely monitor $200,000$-$400,000$ nearby bright stars the so-called TESS Candidate Target List (CTL, \citealt{2018AJStassun,Stassun2019ApJ}) in order to detect periodic transit signals caused by orbiting exoplanets \citep{2015TESSaim}. The telescope began its observations in July $2018$ and continues to observe the sky of each ecliptic hemisphere for one year in a series from $27$-day time spans. During each $27$-day time window, the telescope focuses on one sector covering $24^{\circ}\times 90^{\circ}$ using four cameras with a $2$-minute cadence. In addition to observing the CTL targets, TESS also releases full frame images (FFIs) with longer cadences. The TESS photometric data for CTL targets and FFIs are accessible through the Barbara A. Mikulski Archive for Space Telescopes (MAST) portal\footnote{\url{https://archive.stsci.edu/}}. The TESS telescope was successful in its main aim. According to the NASA exoplanet archive, it has found more than $7000$ candidate exoplanets, of which around $500$ exoplanets have been confirmed. However, its observing strategy including its great photometric accuracy and short cadence is indeed proper for discerning any other periodic or even transient variations in light curves of nearby and bright stars. The sources of such variations can be gravitational microlensing \citep{2023HarrisTESSmicro,2025AJSajadiankalantari}, eclipsing and self-lensing signals in edge-on binary systems \citep{2024ApJSorebella,2024sajadianafshordi}, intrinsic variations or other periodic trends \citep{2023ApJSTESSvariability}, etc. 

In this study, we probe the TESS photometric data taken from seventeen known DWDs to find any periodic signals or trends. We apply the Singular Spectrum Analysis (SSA) technique to de-noise the TESS data for these targets. Then, we evaluate their periodograms which could manifest if there are periodic trends in their light curves. We find these trends in six DWD systems and discuss on their natures by evaluating the False Alarm Probability (FAP), blending effect, and amplitudes of their ellipsoidal variations and Doppler boosting.

\noindent The outline of the paper is as follows: In Section \ref{sec2}, we present a list of seventeen known DWD systems and review their discovered properties. For these systems we will extract their TESS light curves. In Section \ref{sec3}, we first review the SSA formalism and then use it to de-noise the data. By plotting periodograms, we show that six light curves exhibit periodic sinusoidal trends. In Section \ref{sec4}, we focus on these light curves and discuss on natures of their periodic trends. The results and conclusions are reported in Section \ref{sec5}.

\begin{deluxetable*}{c c c c c c c c c c}
\tablecolumns{10}
\centering
\tablewidth{0.95\textwidth}\tabletypesize\footnotesize
\tablecaption{\label{tab1}}
\tablehead{\colhead{$\rm{Name}$}&\colhead{$\rm{TESS}~\rm{ID}$}&\colhead{$\rm{RA}$}&\colhead{$\rm{DEC}$}&\colhead{$T$}&\colhead{Sector}&\colhead{$\tau_{\rm{TESS}}$}&\colhead{$\rm{T}_{\rm{TESS}}$}& $\rm{FWHM}$&\colhead{$\rm{Refs}$}\\
& & $\rm{deg}$&$\rm{deg}$&$\rm{days}$&$\rm{deg}$& $\rm{min}$& $\rm{days}$ & $\rm{days}$ & }
\startdata    
J1717+6757& 219868627 & 17:17:08.50 & +67:57:12.00 & 0.246137& 25,40-41, 47-55 & 10,~30 & 0.246135& 5.525e-5 & (1,2) \\
J1557+2823& 1101592282& 15:57:08.48 &+28:23:36.02 & 0.40741& 51& 10 & 0.793656& 0.02457 & (3)\\
LP400-22    & 2002564035& 22:36:30.10 & +22:32:24.00 & 1.01016&56 & 3.3 & 1.025496& 0.03354 &(4)\\
J1449+1717& 1101113865& 14:49:57.15 & +17:17:29.33 & 0.29075& 51 & 10 & 0.367355& 0.00477 & (5, 6)\\
J2132+0754& 2000073295& 21:32:28.36 & +07:54:28.24 & 0.25056& 55 & 10 & 1.005455& 0.04345 & (3)\\
J2151+1614& 2000525780& 21:51:59.21 & +16:14:48.72 & 0.59152& 55 & 10 &0.196840&  0.001251 & (5)\\
J1104+0918 & 903174194& 11:04:36.74 & +09:18:22.74 & 0.55319& 46, 72 & 10, 3.3 & ~ && (3)\\
J0112+1835 & 611358291& 01:12:10.24 & +18:35:04.10 & 0.14698& 42,43 & 10 & 3.265755& 0.33357& (7)\\
J1249+2626 & 954183245 &12:49:43.57 & +26:26:04.22 & 0.22906& 49 & 10 & ~ & & (5)\\  
NLTT11748 & 434158293 &03:45:16.83 & +17:48:08.71 & 0.23503& 42-44 & 10 & ~ & &(8, 9)\\
CSS 41177  & 840220096 &10:05:59.10 & +22:49:32.26 & 0.116015& 21, 45, 46, 48, 72 &30, 10, 3.3 & ~& &(10, 11)\\
J1112+1117 & 290904838 &11:12:15.82 & +11:17:45.00 & 0.17248& 45, 46, 72& 10, 3.3 & ~ && (12)\\
WD1242-105&156621877 &12:44:52.65 & -10:51:08.90 & 0.11875& 46 & 10 & ~ & & (13)\\
J0923+3028 & 307907698 &09:23:45.60 & +30:28:05.00 & 0.04496& 21, 48 & 30, 10 & &  & (14)\\
J0056-0611 & 610678231&00:56:48.23 & -06:11:41.62 & 0.04337& 43, 70 & 10, 3.3 & 0.189183  & 0.00049& (3)\\
J0755+4800 & 355764113 & 07:55:19.48 & +48:00:34.07& 0.54627& 20, 47 &30, 10  & & &(3)\\ 
J1046-0153 & 902976417 & 10:46:07.88 & -01:53:58.48 & 0.39539& 62, 72 & 3.3 &  & & (3)\\
\enddata
\tablecomments{(1): \citet{Vennes2011J17176757}; (2): \citet{2014MNRASHermesJ17176757}; (3): \citet{2013ApJBrown}; (4): \citet{2006ApJkawka}; (5): \citet{2015ApJGianninas}; (6): \citet{Bell_2017_photometric}; (7): \citet{2012ApJkilic}; (8): \citet{2009AAKawka}; (9): \citet{2010AAKawka}; (10): \citet{2011ApJParsons};  (11): \citet{2014MNRASBours}; (12): \citet{Hermes_2013}; (13): \citet{2015AJDebes}; (14): \citet{Brown_2010_ELMSurvey};    }
\end{deluxetable*}

\section{A list of known Double White Dwarfs}\label{sec2}
For a list contains $67$ known DWDs, we look for their TESS IDs according to their exact coordinates (right ascension and declination) from catalogs of stars extracted from the Full Frame Images (FFIs) reported by the TESS Science Processing Operations Center \texttt{TESS-SPOC} \citep{Jenkins2016,tessspocstars,TESS-SCOPE} and the the Mikulski Archive for Space Telescopes (MAST) catalog \citep{ctlTESS}. Seventeen DWD systems had TESS IDs, and for others there were no recorded light curves. These DWD systems are mentioned in Table \ref{tab1}. After finding the TESS IDs, we extract their TESS light curves using the python package \texttt{lightkurve} which was well developed to analyze any time series data in particular ones taken by NASA's Kepler and TESS missions \citep{2018ascl.soft12013L,2020ascl.soft03001B}.

\noindent In Table \ref{tab1}, the first and second columns report the known names for these DWD systems and their corresponding TESS IDs. Two next columns represent their right ascension and declination. The fifth column is the reported orbital period. The properties of TESS observations from these targets including the number(s) of sector hosting each target, the observing cadence $\tau_{\rm{TESS}}$, and the period of the most-dominant trend in the data $T_{\rm{TESS}}$ are mentioned in the three next columns. The last column mentions the references reporting that target as a binary white dwarfs. We review their known properties in the following.

{\bf J1717$+$6757}:~This system includes an ELM WD with the mass $\simeq 0.18 M_{\odot}$ rotating around a massive and faint WD with the mass $\simeq 0.9 M_{\odot}$ in a close orbit with the orbital period $0.246137$ days. It was first discovered by \citet{Vennes2011J17176757} through spectroscopic measurements and RV variations. \citet{2014MNRASHermesJ17176757} revisited this target again and captured the absorption lines due to nine metals and two eclipsing signals in the photometric data taken by the ULTRACAM camera on the $4.2$m William Herschel Telescope \citep{2007_ULTRACAMWHT}. The TESS telescope observed this target in $2020$ with a $30$-min cadence, $2021$ and $2022$ a $10$-min cadence and has recorded $35,885$ data points. The source ID of this target in the Gaia data is $1637422647712544512$, and its parallax as measured by Gaia is $5.5988$ mas which means that the distance of this target from us is $178.61$ parsec. The RUWE (Re-normalized Unit Weight Error) value reported in the Gaia data for this target is $1.117$ which manifests that the blending effect in the Gaia observations from this target is ignorable.

{\bf J1557$+$2823}:~This is a DWD system which was discovered through the ELM survey \citep{2013ApJBrown}. The orbital period of this system is $0.4074$ days, and the brighter WD has the mass of $0.49 M_{\odot}$ and the mass of its faint companion is $\gtrsim 0.43 M_{\odot}$. \citet{2014_Gianninas} revisited this target and studied the metals abundance in its spectrum. The TESS telescope observed this target with the cadence $10$ minutes in $2022$ and it was inside sector $51$. The number of data points taken by this telescope from this target is $1,506$. The source ID for this target in the Gaia data archive is $1319676603468508544$, and its parallax as reported by Gaia is $4.04837$ mas. Hence, this target should be at the distance $247.132$ parsec from us. The reported RUWE for this target in the Gaia archive is $0.968$ very close to one,which means the luminosity from blending stars in the Gaia observations is ignorable. However, the primary companion is also faint and whose apparent magnitude in $G$-band is $17.848$ mag.  

{\bf LP400$-$22} (or WD 2234$+$222):~This target is known as a runaway DWD system and is moving away from the Galactic center with a high velocity as $830 \rm{km}/s$ \citep{2006ApJkawka,2009AAvennes,2009ApJkilic}. The orbital period of this system is $1.01016$ day, masses of its components are $0.17 M_{\odot}$ and $\gtrsim 0.4 M_{\odot}$, and it was predicted that it is inside the Galactic halo. The TESS telescope observed this target with a $3.3$-min cadence in $2022$ and it was in sector $56$. The number of TESS data points taken from this target is $10,120$. The short cadence and high number of data points are suitable to capture any variations in the photometric data. The Gaia ID of this target is $1874523804732334464$ and its parallax as reported by this telescope is $2.7338$ mas which means its distance from the Earth is $365.7964$ parsec. The reported RUWE value for this target (i.e., $1.0503$) manifests that its blending is ignorable. The apparent magnitude of this target in the Gaia $G$-band is $17.2453$ mag.  

{\bf J1449$+$1717}:~\citet{2015ApJGianninas} discovered this DWD system in the ELM survey observations. This binary system contains two white dwarfs with masses $0.168 M_{\odot}$ and $\geq 0.59 M_{\odot}$ and a short orbital period as $0.29075$ days. The photometric observations from this target was done and no eclipsing signals were recognized \citep{Bell_2017_photometric}. The TESS telescope recorded $1,577$ data points from this target in $2022$ with the observing cadence $10$ minutes. It was inside sector $51$. The source ID for this target in the Gaia data is $1236250127917190272$. This telescope reported its parallax $1.63029$ mas which leads that its distance from the Earth is $613.3863$ parsec. 
Its apparent $G$-band magnitude is $17.7195$ mag.  

{\bf J2132$+$0754}:~A DWD system contains an ELM WD with the mass $0.17 M_{\odot}$ orbiting a massive WD with the mass $\geq 0.95 M_{\odot}$ in a close orbit with period $0.25056$ days \citep{2013ApJBrown}. If this system is edge-on as seen by the observer, self-lensing flare and eclipsing signal likely will be captured in its light curve. The TESS telescope observed this target in $2022$ and during one observing time window from sector $55$ with a $10$-min cadence. This telescope recorded $2,620$ data points. The Gaia source ID for this target is $1740741380258586624$ and its parallax is $0.818788$ mas. Therefore, its distance from us is $1221.31745$ parsec. Its apparent $G$-band magnitude is $18.2546$ mag and it is faint. Its RUWE value ($0.97944045$) tells us its blending effect in its Gaia observations is ignorable.  
 
{\bf J2151$+$1614}:~This DWD system was discovered by \citet{2015ApJGianninas}. Its WDs have masses $0.176M_{\odot}$ and $\geq 0.49 M_{\odot}$ with the orbital period $0.59152$ days. The TESS telescope observed this target during $2022$ with a $10$-min observing cadence. It was in sector $55$ and the number of TESS data points taken from this target is $2,601$. The Gaia source ID of this target is $1772771670097796864$. The reported parallax for this target by Gaia is $2.55616$ mas. Hence, its distance from us is $391.2119$ parsec. The RUWE reported for this target is $1.0428$. Its apparent $G$-band magnitude is $16.8989$ mag and it is brighter than two previous target.  

{\bf J1104$+$0918}:~This DWD system was first announced by \citet{2013ApJBrown}. The masses of WDs in this binary system are $0.46 M_{\odot}$ and $\geq 0.55 M_{\odot}$ and their orbital period is $0.55319$ days. Hence, two WDs in this system are more massive than $0.4$ solar mass. TESS detected this target in $2021$ and $2023$ and while observing sectors $46$ and $72$ with the observing cadences $10$ and $3.3$ minutes, respectively. The total number of TESS data points from this target is $12,148$. Since, there were no dominant and periodic trends in its TESS data, we put aside this target.  

{\bf  J0112$+$1835}:~It is a close DWD system (including one ELM WD and a massive one) with a short orbital period $0.15$ days \citep{2012ApJkilic}. Constraints on the mass of the faint companion and the inclination angle of its orbital plane were done through measuring ellipsoidal variations \citep{Hermes_2014_ellipsiodal}. They concluded that the inclination angle of the orbital plane is $70$ degrees. TESS observed this target in $2021$ with a $10$-min cadence. It was inside sectors $42$ and $43$ so that the number of TESS data points recorded from this target was $5,317$. There is an obvious long-period and large amplitude ($0.1$ in the normalized flux) trend in its light curve with the period $3.267$ days. This period is very longer than its orbital period. The TESS $T$-band magnitude for this target is $17$ mag, so its data have high noises and we do not study this target more.    

{\bf J1249$+$2626}:~This is a DWD system which contains an ELM WD with the mass $0.16 M_{\odot}$ orbiting another WD with mass $\geq 0.35M_{\odot}$ in a close orbit. Their orbital period is $0.22906$ days \citep{2015ApJGianninas}. The TESS telescope detected this target in $2022$ with a $10$-min cadence and recorded $2,668$ data points. It was inside sector $49$. 

{\bf NLTT11748}:~This is an eclipsing DWD system which was first discovered by \citet{2009AAKawka,2010AAKawka} through RV measurements. This system includes one ELM WD with mass $0.17 M_{\odot}$ rotating in a close orbit around a massive WD with mass $0.75 M_{\odot}$  and the orbital period $0.2350$ days. \citet{2014ApJKaplan} observed this target with a short cadence and discerned its eclipsing signals. TESS recored $8,743$ data points from this target in $2021$ with a $10$-min observing cadence. This target was in three sectors $42$, $43$, and $44$. Although this system is eclipsing, but these signals were not recovered in the TESS data.     

{\bf CSS 41177}:~This is a double helium white dwarf binary with deep eclipsing signals. The masses of its WDs are $0.38 M_{\odot}$ and $0.32 M_{\odot}$ with a short orbital period $0.116015$ days \citep{2011ApJParsons,2014MNRASBours}. The TESS telescope detected this target several times in $2020$-$2023$ and with different cadences $30$, $10$, and $3.3$ minutes while this target was inside five sectors $21$, $45$, $46$, $48$, and $72$. But, its first set of its data (related to sector $21$) was only available. In this set of data periodic signals or trends were not realizable. 

{\bf J1112$+$1117}:~\citet{Hermes_2013} first announced the discovery of this eclipsing DWD system. The TESS telescope detected this target in $2021$ and $2023$ with the cadence $10$ and $3.3$ minutes, respectively. This target was inside three sectors $45$, $46$, and $72$. For this target, only $3,238$ data points were available which were not enough to recover its eclipsing signals or periodic trends.    

{\bf WD1242$-$105}:~This is a DWD system which was first reported in \citet{2015AJDebes}. This system is at the distance $35$ parsec away from us, and contains two WDs with the total mass $0.95$ solar mass with a short orbital period $2.85$ hours. The TESS telescope detected this target in $2021$ with a $10$-min cadence. It was inside sector $46$. During this observation, $3,184$ data points were recorded from it. There was no periodic signals or trends in the TESS data.  

{\bf J0923$+$3028}:~This DWD system contains two WDs with masses $0.279 M_{\odot}$ and $\gtrsim 0.37M_{\odot}$ rotating in a close orbit with the period $0.04495$ days \citep{Brown_2010_ELMSurvey}. The TESS telescope detected this target in $2020$ and $2022$ with cadences $30$ and $10$ minutes, respectively. The target was inside sectors $21$, and $48$. The number of TESS data points taken from this target is $4,048$. For this system the orbital period is too short so that finding any periodic trends or signals with periods similar to the orbital period is hard and needs dense photometric observations. We de-noised its TESS data but could not find any periodic trends or signals.  

{\bf J0056$-$0611}:~This is a DWD system and contains an ELM WD with mass $0.17 M_{\odot}$ rotating a more massive WD with the mass $\gtrsim 0.46 M_{\odot}$ in a very close orbit with the orbital period $0.04338$ days. The TESS telescope detected this target in $2021$ and $2023$ with cadences $10$ and $3.3$ minutes, respectively. This target was inside two sectors $43$ and $70$. The number of TESS data points from this target is $11,646$.  

{\bf J0755$+$4800}:~This DWD system contains two WDs with masses $0.261 M_{\odot}$ and $0.409 M_{\odot}$ with the orbital period $0.54627$ days \citep{2013ApJBrown}. TESS detected this target in $2019$ and $2021$ with cadences $30$ and $10$ minutes. That was inside two sectors $20$ and $47$. The total number of data points from this target taken by TESS is $4,404$.    

{\bf J1046$-$0153}:~This is a DWD system which was first discovered from the ELM survey observations \citep{2013ApJBrown}. WDs in this system have masses $0.375 M_{\odot}$ and $\geq 0.19 M_{\odot}$ and rotate in a close orbit with the orbital period $0.39539$ days. Such low mass WDs in close orbits could make eclipsing signals, if their orbital plane is edge-on as seen by the observer. The TESS telescope detected this target in $2023$ with a $3.3$-minute cadence. In fact, the target was inside two sectors $62$ and $72$ and the number of the TESS data points from this target was $16,904$. 

\noindent In the next section we first de-noise the TESS light curves and then search for any periodic trends or signals in these light curves.

\section{Analyzing the TESS DWD light curves}\label{sec3}
We first de-noise the time-series data taken by TESS from targets listed in Table \ref{tab1} based on the SSA technique. This method is briefly explained in Subsection \ref{sub1}. To extract any periodic and regular signals or trends from these TESS light curves we plot their periodograms as explained in Subsection \ref{sub2}. If there are periodic features in time-series data whose periods are either close to the orbital period or proper fractions of it, the detected signals potentially could be due to the orbital motion (e.g., eclipsing, self-lensing, ellipsoidal variations, Doppler boosting) and could give us some information about the physical parameters of components. We test the significance of detected trends based on FAP values as explained in Subsection \ref{subnew}. Finally, we fold the data into the time intervals equal to the periods of the most-dominant feature (with the maximum power in the periodograms) to realize trends and signals. In Subsection \ref{sub3}, we explain all results. 

\subsection{Singular Spectrum Analysis method}\label{sub1}
Singular Spectrum Analysis (SSA) is a powerful non-parametric technique used to make spectra of all periodic signals in time-series data. SSA consists of two complementary stages. The first stage involves decomposing a time series into a sum of principal components by embedding it into a trajectory matrix and performing singular value decomposition (SVD). The second stage includes reconstructing the time series data by grouping and then diagonal averaging. The only parameter involved in this technique is the window length $L$ which represents the number of components we aim to extract from the data. $L$ should be less than $N/2$, where $N$ is the number of data points in the considered time series.\\
Let's consider a time series as $Y_{\rm N}=\big(y_{1},~y_{2},~y_{3}, ...,~y_{\rm N}\big)$. SSA first makes a trajectory matrix from this time series as: 
\begin{eqnarray}
\textbf{X}=
\begin{pmatrix}
y_{1} ~& y_{2} ~& y_{3} ~& \cdots ~& y_{K} \\
y_{2}~& y_{3} ~& y_{4} ~& \cdots ~& y_{K+1} \\
\vdots ~& \vdots ~& \vdots ~& \ddots ~& \vdots \\
y_{L} ~~& y_{L+1} ~~& y_{L+2} ~& \cdots ~~& y_{N}\\
\end{pmatrix},
\end{eqnarray}
where $K=N-L+1$. This matrix is the $L\times K$ Hankel matrix where $X_{ij}$s are equal when $i+j=\rm{const}$. The columns of this matrix are the so-called lagged vectors with the size $L$. We can make a square matrix as $\textbf{S}=\textbf{X} \textbf{X}^{T}$ where $\textbf{X}^{T}$ is the transpose of the trajectory matrix. For this matrix generally there are $L$ eigenvalues $\lambda_{1},~\lambda_{2},~\lambda_{3},~...,~\lambda_{L}$, whose corresponding orthonormal eigenvectors are denoted by $U_{1},~U_{2},~U_{3}, ..., U_{L}$. 
	
\noindent We define $d$, the rank of the trajectory matrix, as the number of non-zero eigenvalues $d=\rm{max}\{i,~\rm{if}~\lambda_{i}>0\}$. After making the vectors $V_{i}=\textbf{X}^{T} U_{i}/\sqrt{\lambda_{i}}$ where $i\in 1, 2, ..., d$, we then perform the singular value decomposition of the trajectory matrix as given by:
\begin{eqnarray}
\textbf{X}={\bf U}~{\bf \Sigma}~{\bf V^{T}}= \textbf{X}_{1}+ \textbf{X}_{2}+... \textbf{X}_{L}, 
\end{eqnarray}
where ${\bf \Sigma}$ is $L\times K$ matrix which its diagonal elements are $\Sigma_{ii}=\sqrt{\lambda_{i}}$ and its non-diagonal elements are zero. ${\bf U}$ is a $L\times L$ matrix which its columns are $U_{i}$s, and $\textbf{V}$ is a $K\times K$ matrix whose columns are $V_{i}$s. We note that $\textbf{X}_{i}=\sqrt{\lambda_{i}}~U_{i}~V_{i}^{T}$ are the elementary matrices with one eigenvalue and the triplet $(\lambda_{i}, U_{i}, V_{i}^{T})$ is the so-called $i$th eigentriplet. 
		
In the grouping step, we set ${\bf X}_{i}$s into $m$ subgroups as $\textbf{X}_{{I}_{1}}$,~$\textbf{X}_{{I}_{2}}$,~$...$,~$\textbf{X}_{{I}_{m}}$, where every $\textbf{X}_{{I}_{i}}$ is itself the summation of several elementary matrices. In this step we may remove some elementary matrices to de-noise data. Finally, we have a matrix $\textbf{Z}=\textbf{X}_{{I}_{1}}+\textbf{X}_{{I}_{2}}+...+\textbf{X}_{{I}_{m}}$ and should extract its corresponding time-series data by diagonal averaging for each of $\textbf{X}_{{I}_{i}}$ where $i\in 1,~2,~...,~m$. The diagonal averaging means we first make their corresponding Hankel matrix $\widetilde{\textbf{X}}_{{I}_{i}}$ (the s-called Hankelization process) and then in the same way as making the trajectory matrix (in the embedding step) we convert that resulted Hankel matrix (i.e., $\widetilde{\textbf{Z}}=\widetilde{\textbf{X}}_{{I}_{1}}+\widetilde{\textbf{X}}_{{I}_{2}}+... +\widetilde{\textbf{X}}_{{I}_{m}}$) to time series data.

\subsection{periodogram}\label{sub2}	
To extract periodic signals and trends from DWD light curves, we plot their periodograms using two techniques: (i) Box Least Square (BLS, \citealt{2002AABLSperiod}) and (ii) Lomb-Scargle (LS, \citealt{1976Lomb,1982Scargle}). The BLS method is optimized for exoplanetary transit signals. It divides a time series into several boxes by considering four parameters: period, duration, depth, and a reference time. For a given period, the best-fitted values for other parameters are determined by maximizing the likelihood function. Hence, this method is perfectly suitable for finding eclipsing signals in DWD light curves. The LS technique, on the other hand, finds periodic signals using uneven temporal sampling from data. In this method, Fourier-like waves are fitted to data by considering a wide range of frequencies and then a Fourier power spectrum is generated \citep{LombScargel}. We note that eclipsing and lensing signals do not have sinusoidal shapes. Therefore, LS periodograms are very suitable for finding sinusoidal-like variations in data, which could be due to ellipsoidal variations or Doppler boosting in close and interacting binaries or intrinsic changes. 

\subsection{Significance Test}\label{subnew}
To evaluate the significance of detected periodic trends, we calculate FAP values by generating pure noisy normalized fluxes for each given time series. Each pure noisy normalized flux in a given time is given by a normal distribution with the width equals to the reported photometric error on that time. For each target we make $\sim200-2000$ pure noisy models and then plot their periodograms in the same frequency interval. We finally evaluate the FAP value as the per cent of simulated pure noisy light curves whose maximum periodogram powers are larger than the maximum of the target's periodogram ($\rm{Power}_{\rm{max}}$) \citep[see, e.g.,  ][]{2008Frescura}. However, there are some other sources for false alarms, such as blending effect due to a regular variable star, i.e., pulsating, eclipsing stars \citep{2011Morton}. Here, we evaluate the FAP based on only noises. For one of targets we investigate the blending effect using the Gaia catalog. Generally, if the FAP value is $\gtrsim 10\%$, the detected periodic trend is not reliable and likely an artifact of the data. When $\rm{FAP}\lesssim1\%$ the detected trend is very secure.

\begin{figure*}
\centering
\includegraphics[width=0.48\textwidth, height=0.2\textheight]{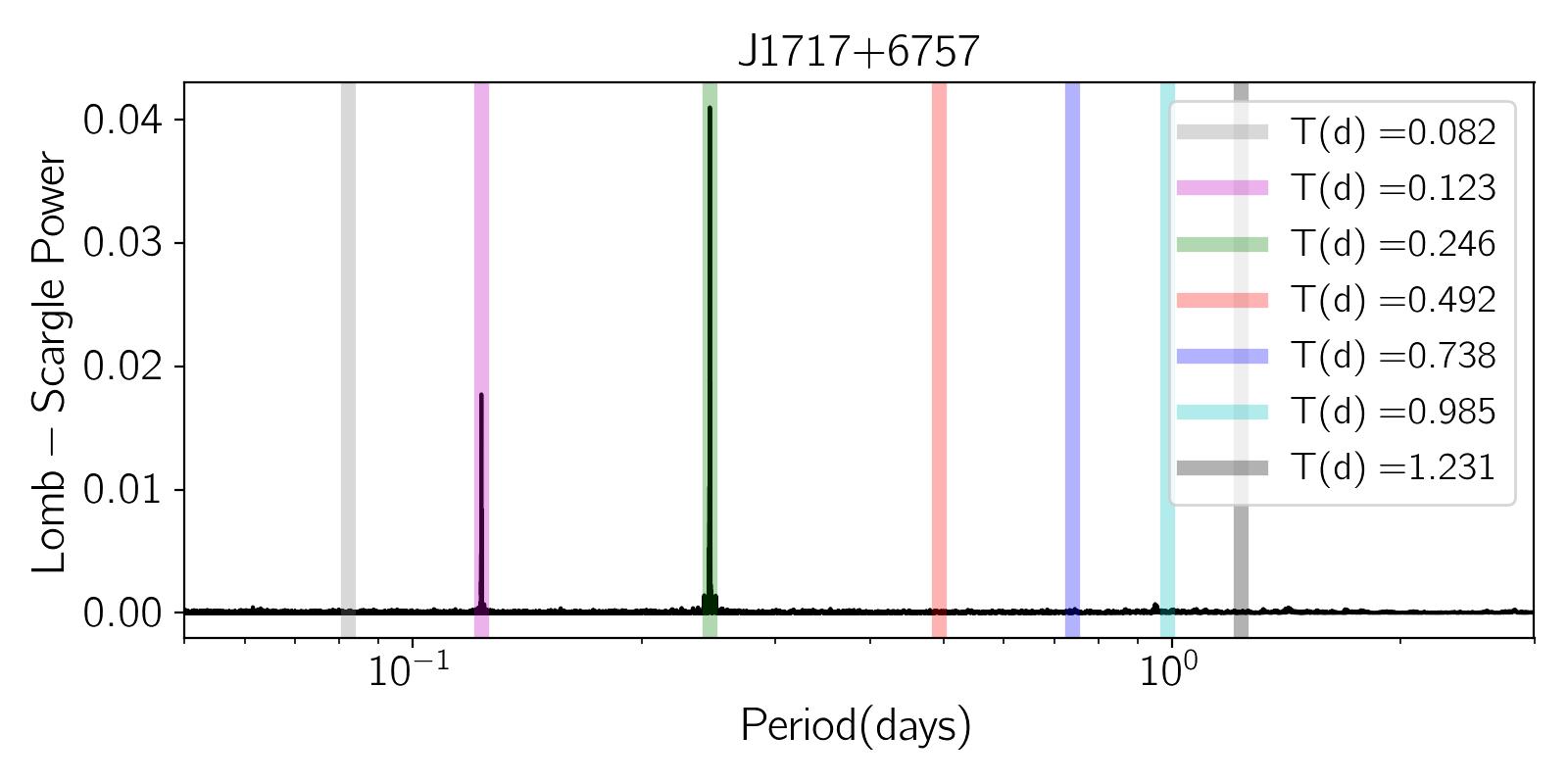}
\includegraphics[width=0.48\textwidth, height=0.2\textheight]{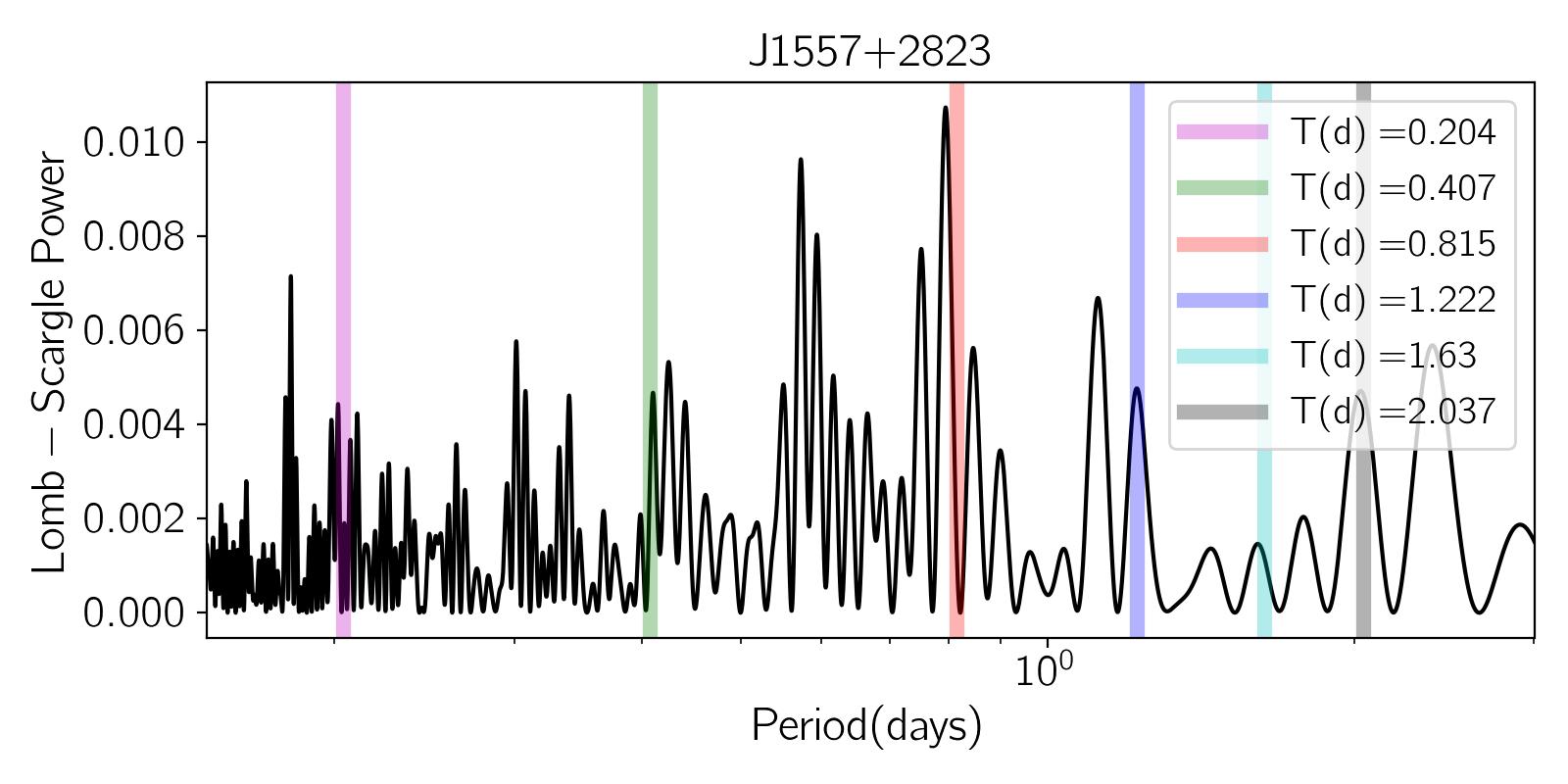}
\includegraphics[width=0.48\textwidth, height=0.2\textheight]{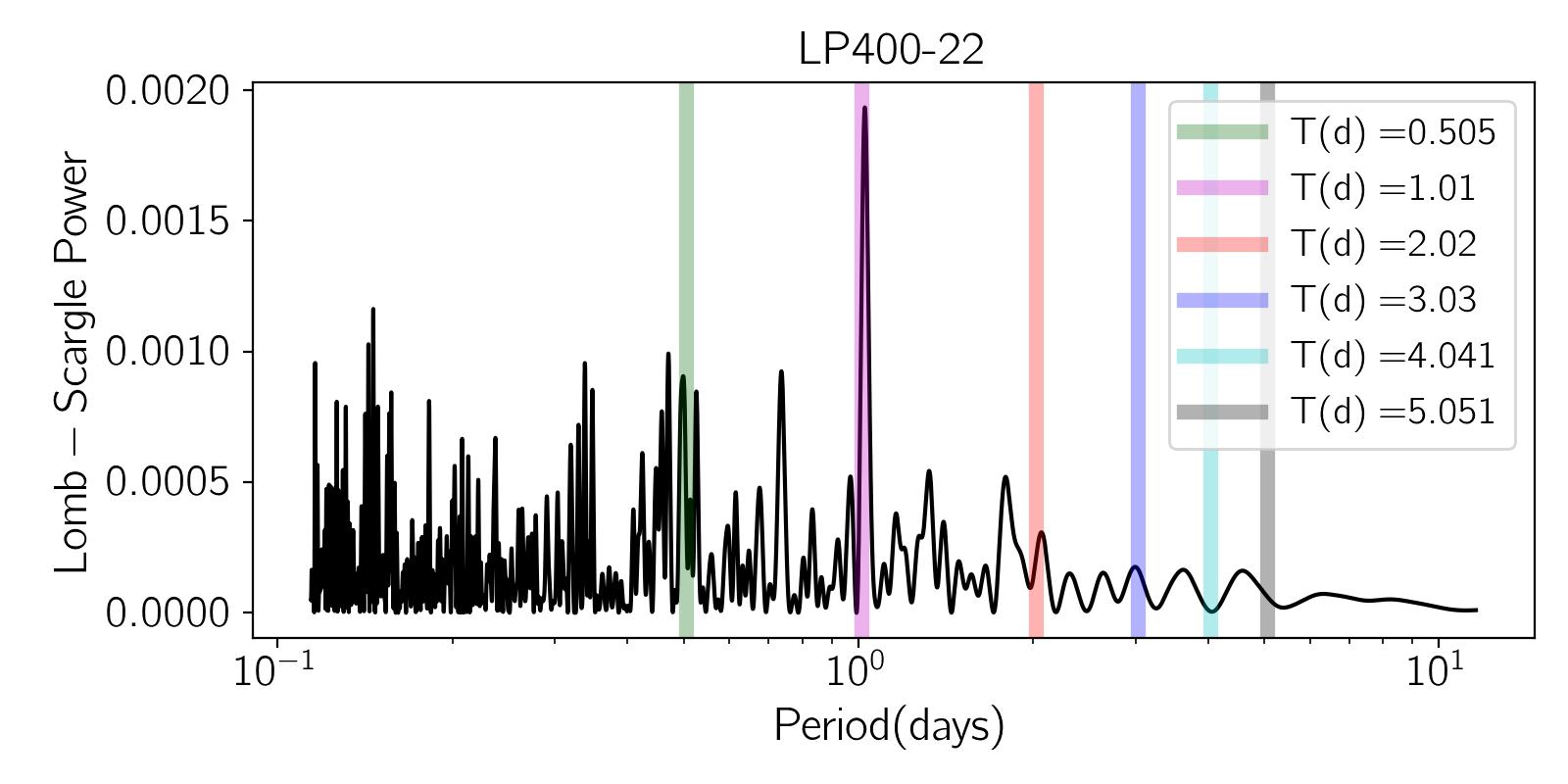}
\includegraphics[width=0.48\textwidth, height=0.2\textheight]{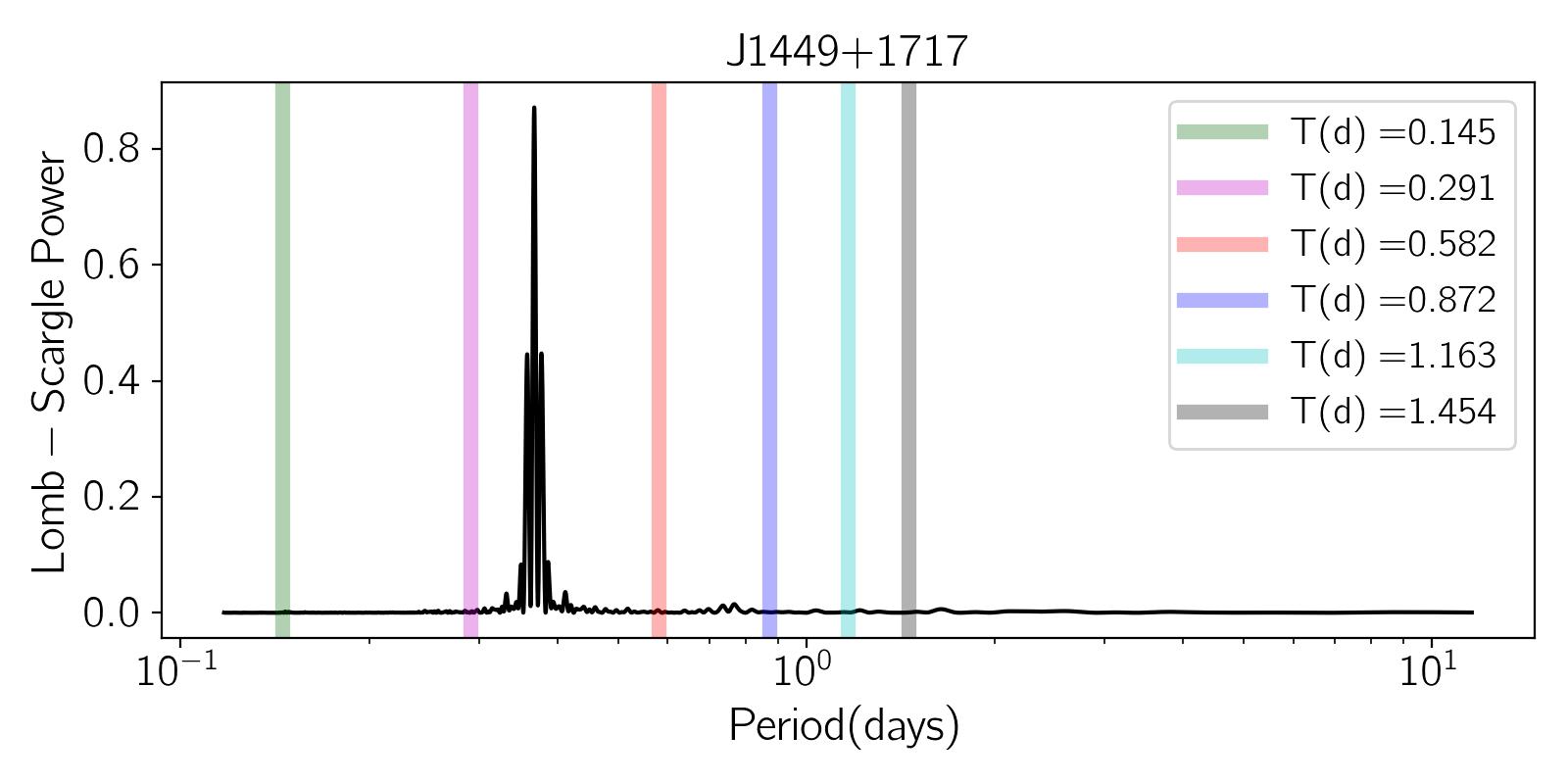}
\includegraphics[width=0.48\textwidth, height=0.2\textheight]{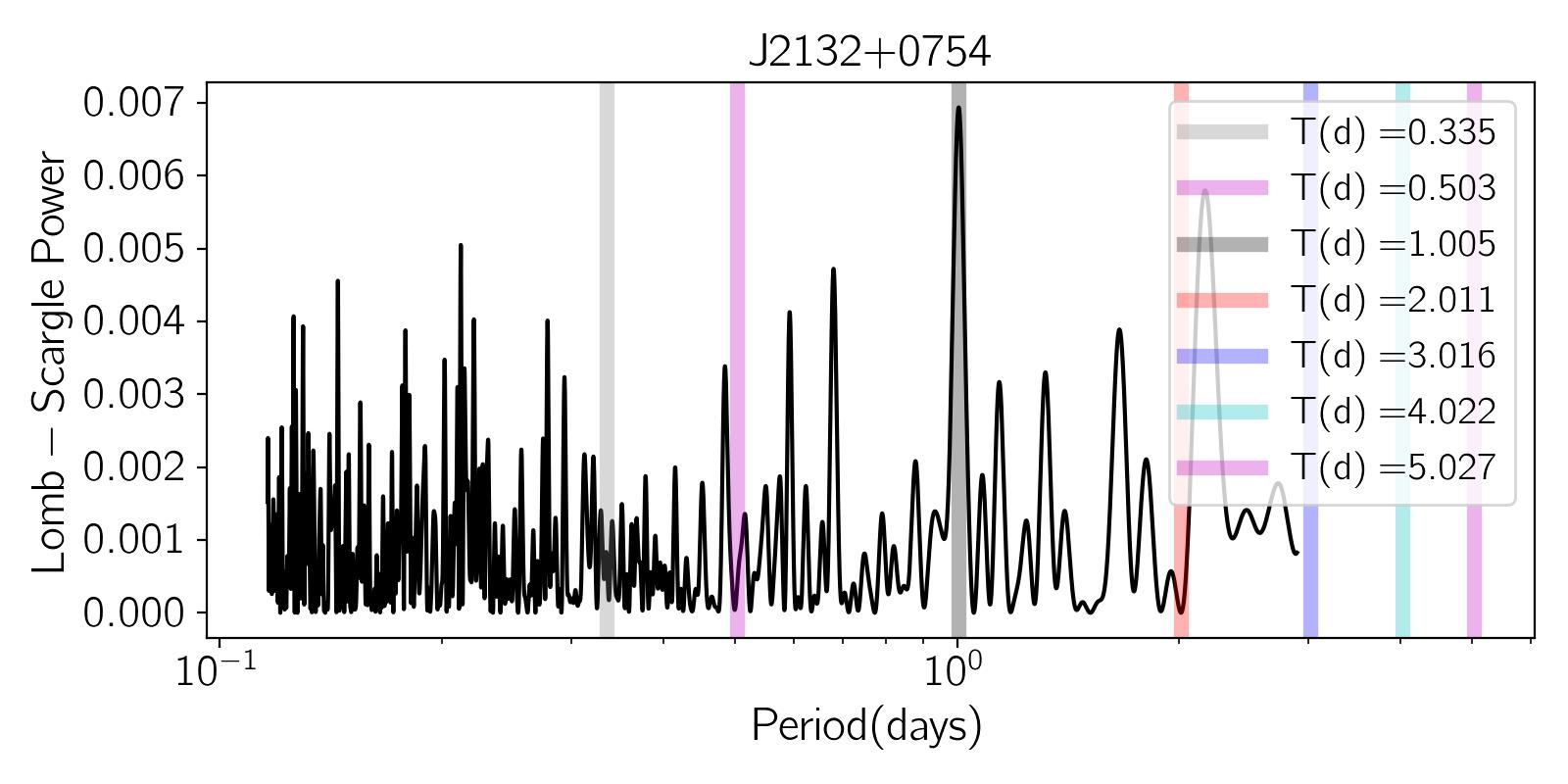}
\includegraphics[width=0.48\textwidth, height=0.2\textheight]{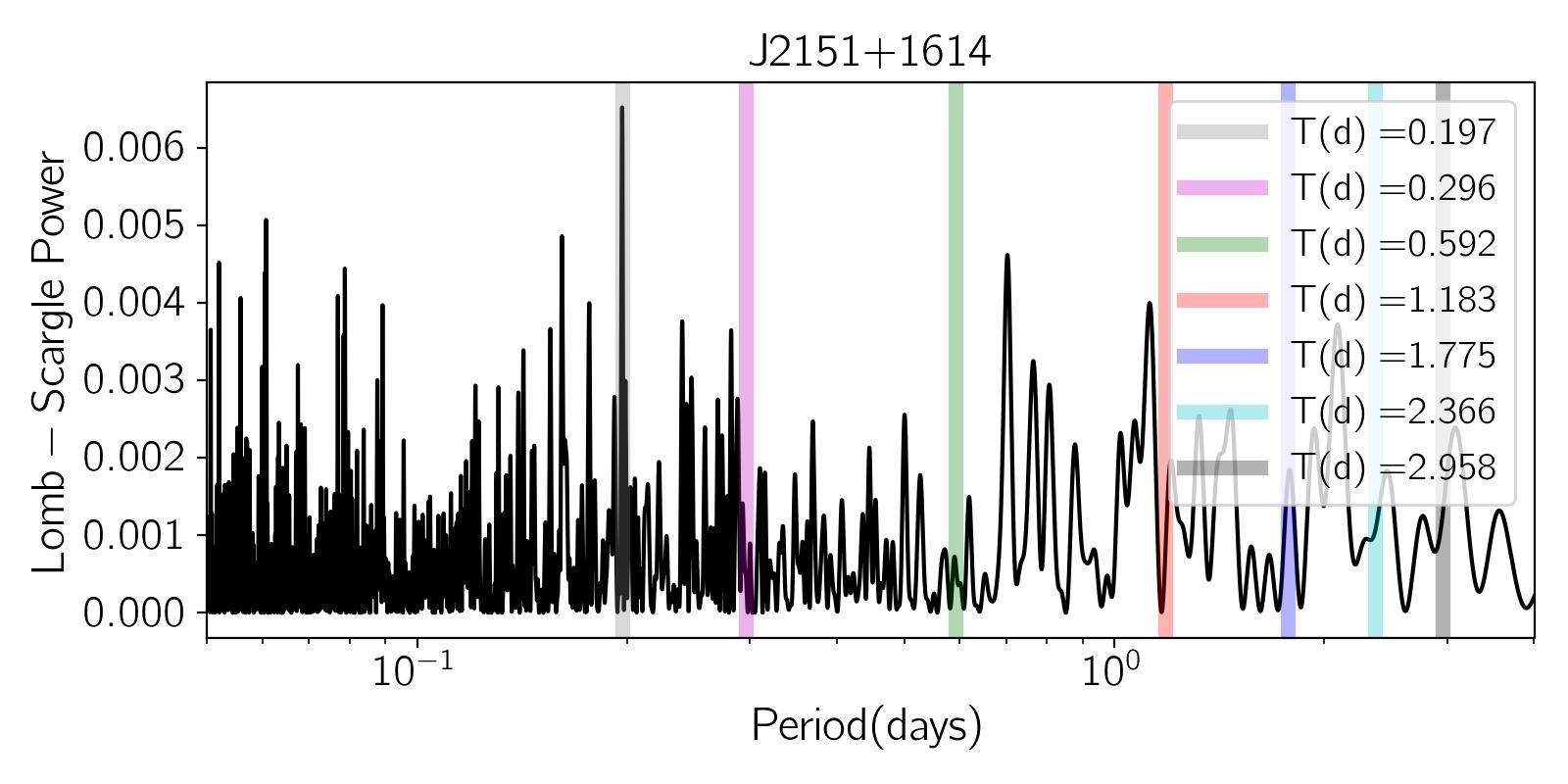}
\caption{The LS periodograms related to the TESS light curves of six known DWD systems. The name of each target is reported at the top of each panel. The thick colored and vertical lines specify the orbital period, its half, double, triple, etc. These lines are depicted to compare the orbital period and its proper fractions/multiples with the period of their most-dominant signal with the maximum LS power.}\label{fig1}
\end{figure*}

\begin{figure*}
    \centering
    \includegraphics[width=0.48\textwidth]{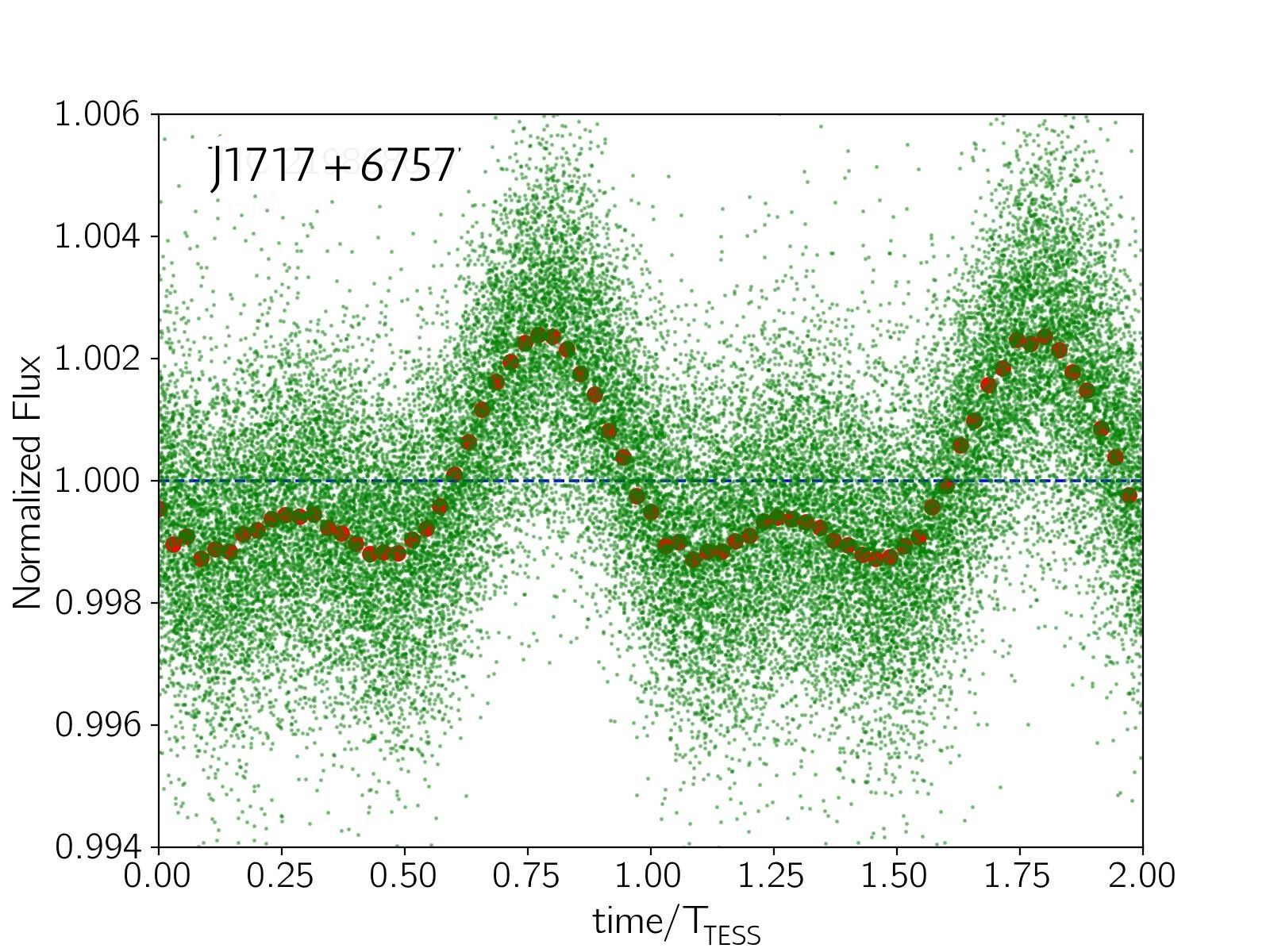}
    \includegraphics[width=0.48\textwidth]{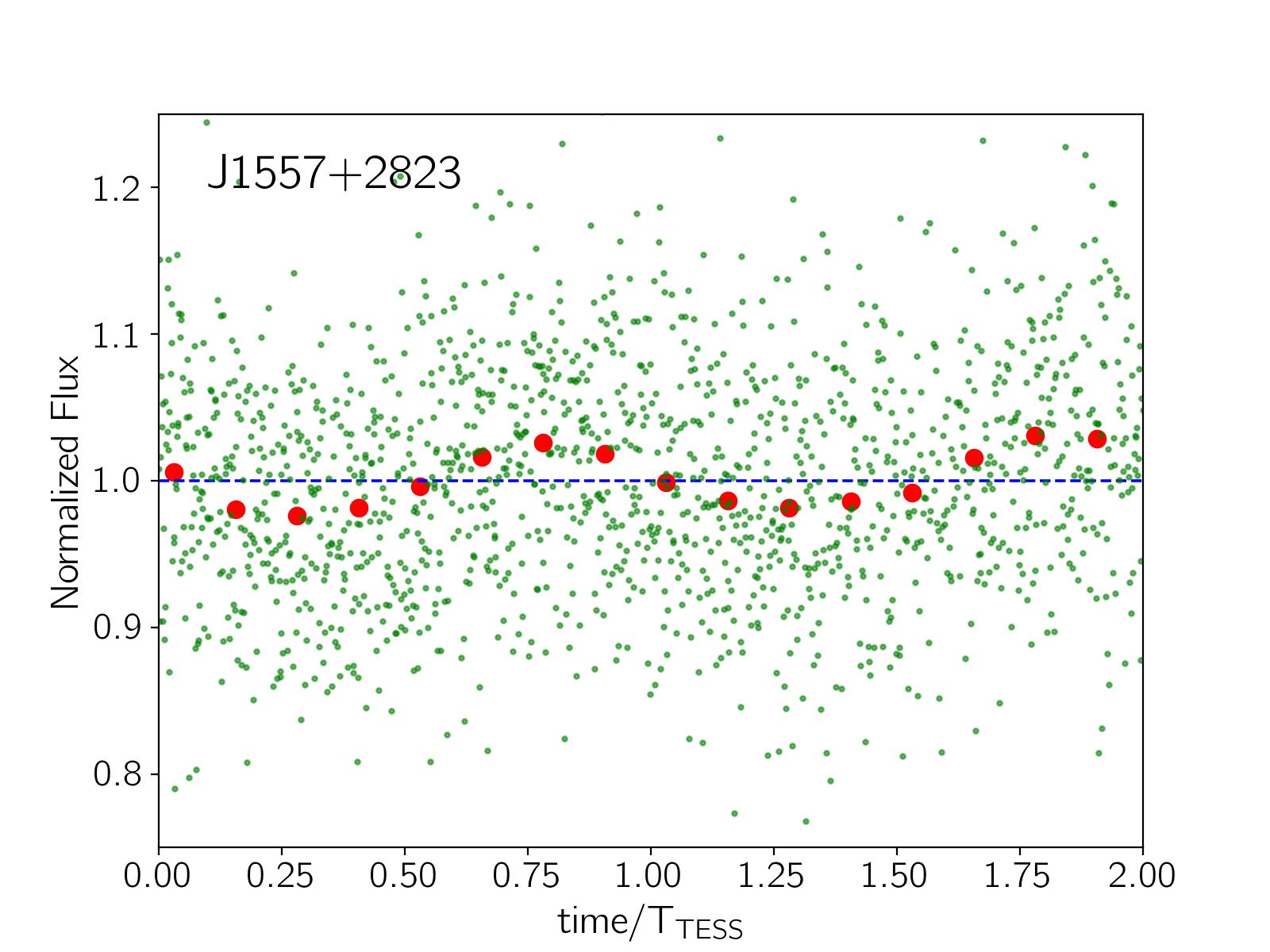}
    \includegraphics[width=0.48\textwidth]{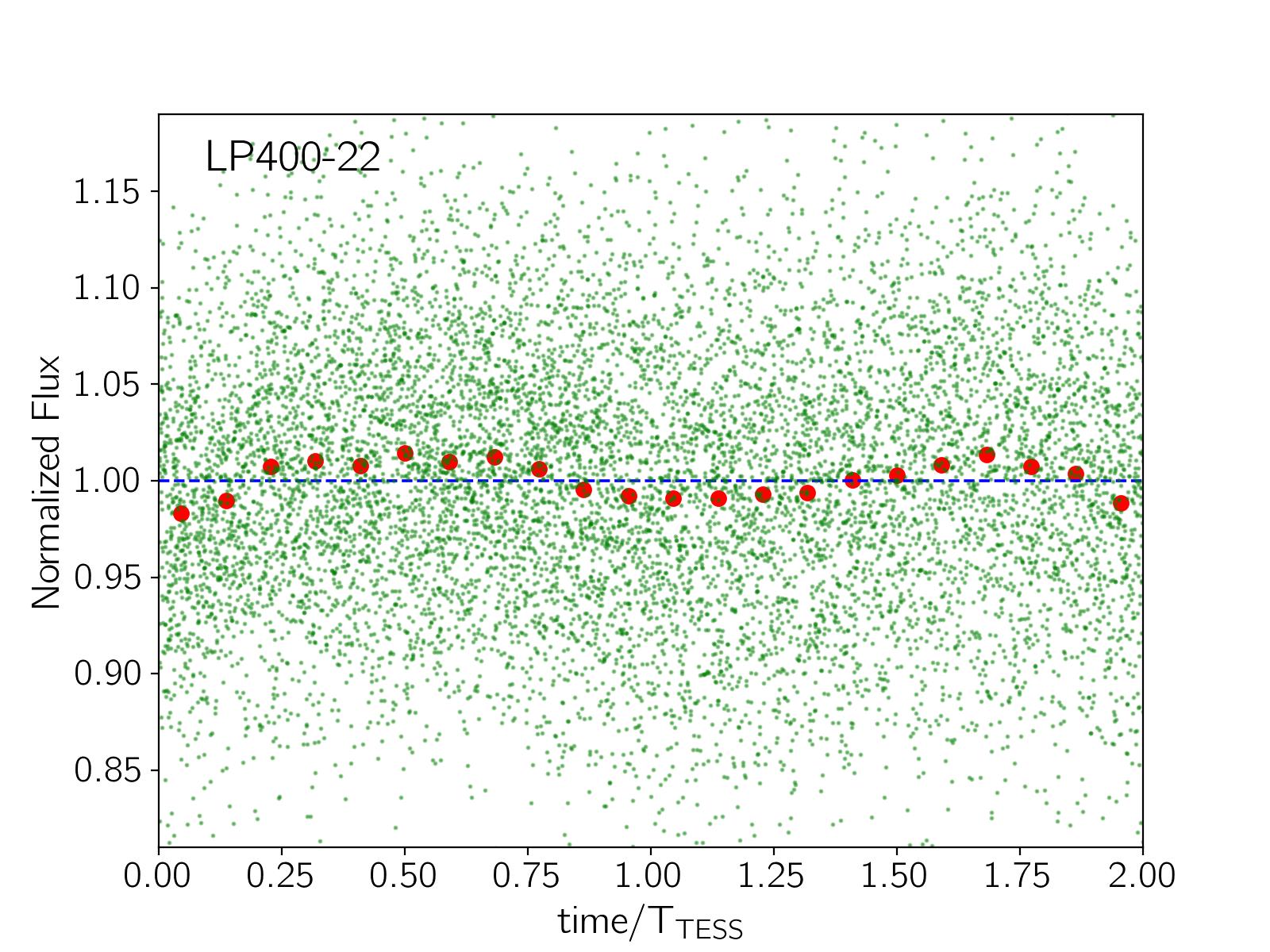}
    \includegraphics[width=0.48\textwidth]{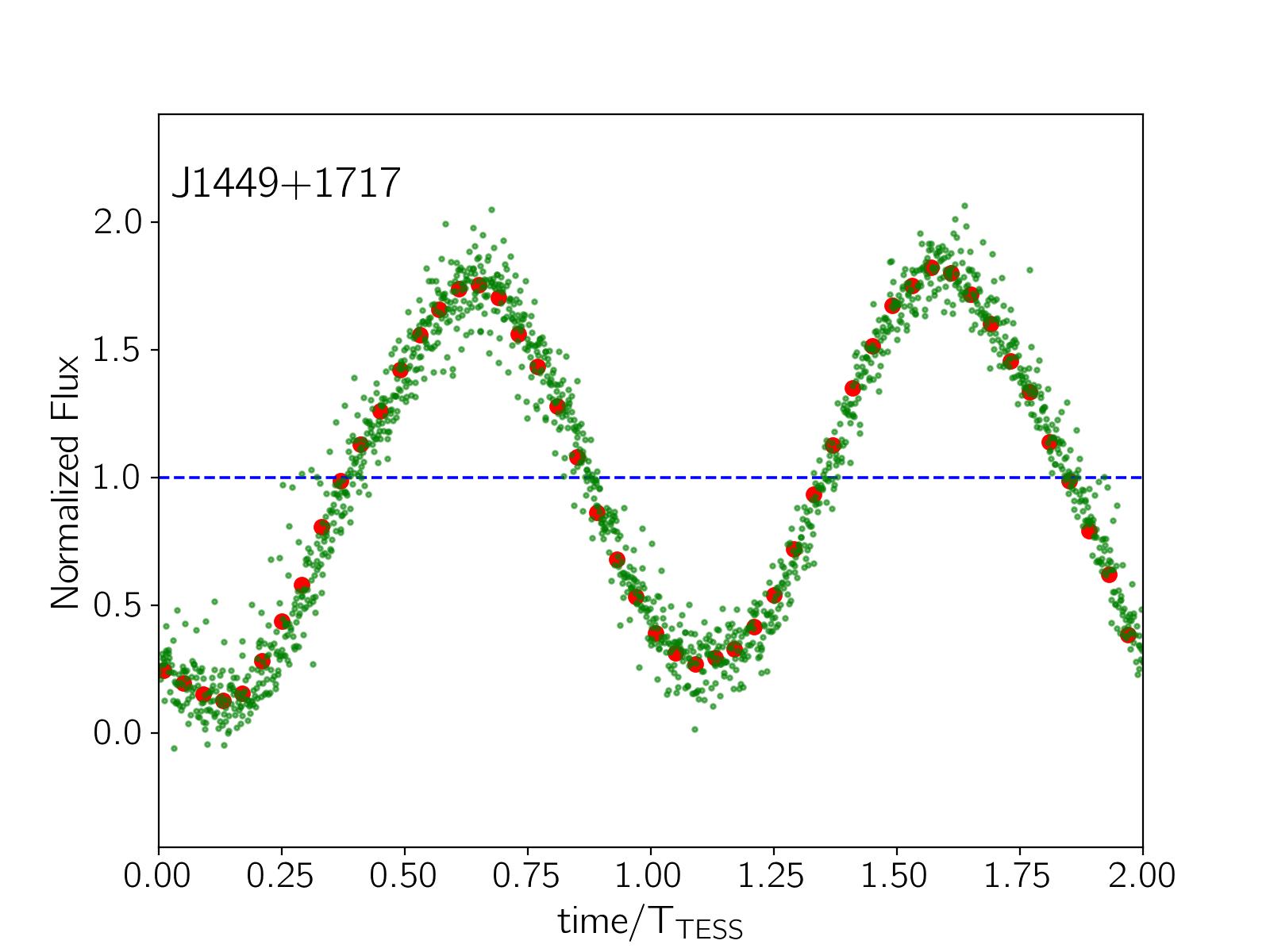}
    \includegraphics[width=0.48\textwidth]{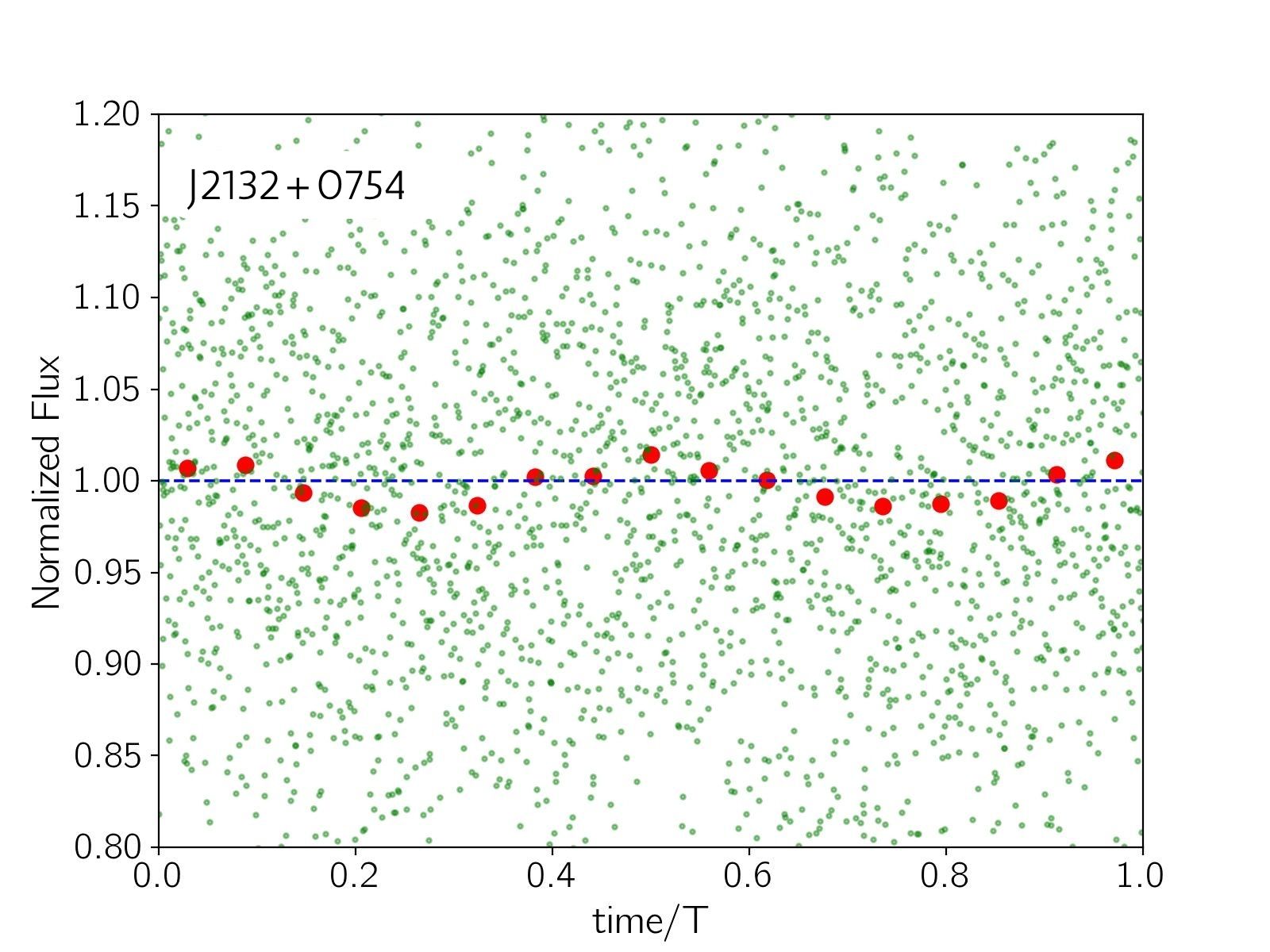}
    \includegraphics[width=0.48\textwidth]{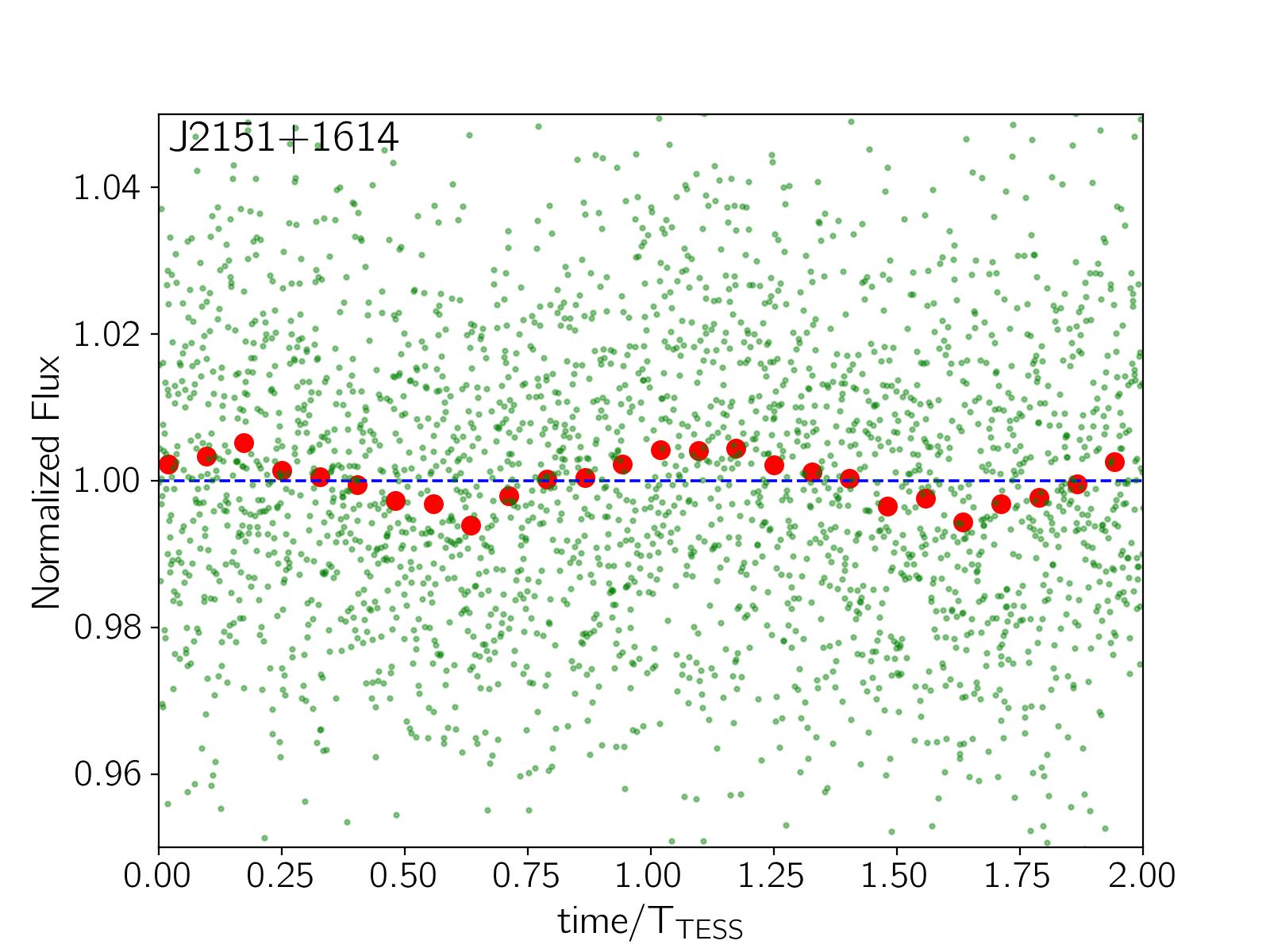}
    \caption{The TESS folded light curves for six DWD systems whose periodograms are displayed in Figure \ref{fig1}. The name of each system is mentioned inside the panel. The binned data are represented by red circles. To better discern the periodic trends in data, we have included horizontal blue lines indicating constant flux in each panel.}\label{fig2}
\end{figure*}    

\begin{deluxetable}{c c c c c c c}
\tablecolumns{7}
\centering
\tablewidth{0.49\textwidth}\tabletypesize\footnotesize
\tablecaption{\label{tab1b}}
\tablehead{\colhead{$\rm{Name}$}&\colhead{$N$}&\colhead{$\sigma^{2}_{\rm{N}}$}&\colhead{$\overline{\delta_{\rm{F}}}$}&$\Delta_{\rm n}$&\colhead{$\rm{Power}_{\rm{max}}$}&\colhead{$\rm{FAP}$}\\
& & &  & & & $\%$ }
\startdata    
J1717+6757 & 35885 & 1.755e-5 & 0.004 & 0.003 & 0.07 &  $<0.33$ \\
J1557+2823 & 1506 & 0.042 & 0.219 & 0.031&0.011 & 14.83 \\
LP400-22  & 10120& 0.071 & 0.295 & 0.017& 0.002 & 3.68 \\
J1449+1717 & 1577 & 0.351 & 0.195 & 0.874 & 0.871 &  $<0.08$\\
J2132+0754 & 2620 & 0.098 & 0.329 & 0.017 &  0.007 & 8.82\\
J2151+1614 & 2601 & 0.002 & 0.054 & 0.006 & 0.007 & 36.47\\
    \enddata
    \tablecomments{$\sigma^{2}_{\rm N}$ and $\overline{\delta_{\rm{F}}}$ are the variance and the average of photometric errors in normalized fluxes, respectively. $\Delta_{\rm n}$ is the amplitude of most-dominant trends in data and in the unit of normalized flux. $\rm{Power}_{\rm{max}}$ is the maximum power of LS periodograms.}
\end{deluxetable}

\subsection{Results}\label{sub3}
For the DWD targets mentioned in Table \ref{tab1}, we first plot their BLS and LS periodograms by considering wide ranges of periods. We did not find any eclipsing or lensing signals with periods equal to their orbital periods in BLS periodograms. This could be due to the long cadence of the TESS observations for FFI's targets (all DWD systems in our list were extracted from the TESS FFIs) compared to their duration. For the first six targets mentioned in Table \ref{tab1}, there were dominant and discernible periodic (sinusoidal-like) trends in their LS periodograms. The LS periodograms for these targets are shown in different panels of Figure \ref{fig1}. In each panel, the LS power is shown versus the period with thick black lines. Additionally, each panel includes thick colored and vertical lines that show the orbital period and its proper fractions or multiples for comparison with the period of the most-dominant trend. The period of the most-dominant feature in each light curve $T_{\rm TESS}$ is reported in the eighth column of Table \ref{tab1}. 

For the six DWD targets whose periodograms are represented in Figure \ref{fig1}, the folded data in time intervals of either one or twice $T_{\rm{TESS}}$ are shown in different panels of Figure \ref{fig2} with green data points. In these plots, red circles represent binned data that clearly show periodic patterns and shapes of their most-dominant features. To better illustrate variations in time-series data, we show the constant flux with dashed blue lines in these panels. The name of each target is mentioned inside each panel.

We also report some statistical parameters about these six light curves and their detected trends in Table \ref{tab1b}. In this table, $N$ is the number of the TESS data taken from each target. The third and forth columns of this table report $\sigma_{\rm N}^{2}$ and $\overline{\delta_{\rm F}}$ which are the variance and the average of photometric errors of data in the normalized flux, respectively. Also $\Delta_{\rm n}$ is the amplitude of the most-dominant trend in the normalized flux reported in the fifth column of this table. The amount of maximum power in their LS periodograms $\rm{Power}_{\rm{max}}$ and FAP values can be found in two last columns of this table.

For the first DWD in our list, J1717$+$6757, the largest LS power occurs at a period of $0.246135$ days, which is very close to the reported orbital period with the difference $\simeq 2e-6\%$ which is smaller than the FWHM of the peak of its periodogram. Its folded light curve is shown in the first panel of Figure \ref{fig2}. This trend in the photometric data of this target was also displayed in Figure (3) of \citet{Vennes2011J17176757}. For this target, the average of photometric errors is $0.004$ in the normalized flux unit, as reported in Table \ref{tab1b}. Although the amplitude of its periodic trend is $\sim 0.003$, its FAP value is less than $0.33\%$ (the number of simulated pure noisy data sets was 301). Hence, the periodic trend in its data is not due to noises and has an orbital origin.

The next target, J1557$+$2823 with the TESS ID $1101592282$, has a crowded periodogram that shows different noises in its light curve. According to its periodogram, the most-dominant trend in its data occurs with a period very close to twice its orbital period, i.e. $T_{\rm{TESS}}=0.793656$ days which is $1.95~T$. We de-noise its data using the SSA technique. The folded and de-noised data in the time interval $[0,~T_{\rm{TESS}}]$ is represented in the second panel of Figure \ref{fig2}. A sinusoidal-like trend can be inferred from its binned data points (red circles). We also evaluate the FAP value by simulating pure noisy data sets for this target as $14.83\%$. Hence, this trend is weak and could be due to noises.

The folded light curve and LS periodogram for the third target, LP400$-$22, are shown in the third panels of Figures \ref{fig2} and \ref{fig1}. The most-dominant periodic feature in its data happens at the period $T_{\rm{TESS}}=1.025496$ days, which is close to the reported orbital period for this target with the difference $\simeq 0.015$ days (less than the FWHM of the peak in the periodogram). The variation has a sinusoidal shape and repeats twice in $2 T_{\rm{TESS}}$. Therefore, this variation has a period equal to the orbital period and could be due to Doppler boosting. The average photometric errors of its TESS data is $\sim 0.3$ in the normalized flux, but on the other hand the number of data points recorded from this target is high. Hence, FAP resulting from pure noisy data points for this target is $3.68\%$ which is low enough to keep it for future study.

The light curve of the fourth target, J1449$+$1717, exhibits a powerful periodic feature with a period of $0.367355$ days and an amplitude of $0.874$ in the normalized flux (or $\sim 0.75$ mag in the apparent magnitude) as shown in the fourth panel of Figure \ref{fig2}. This point can be identified from its periodogram (the fourth panel of Figure \ref{fig1}). This variation is a regular sinusoidal-like feature and its period is not a multiple or proper fraction of the orbital period reported for this target. Hence, this periodic trend could be intrinsic or due to blending effect. We also evaluate the FAP for this target based on simulating pure noisy data pints which was less than $0.08\%$. According to the folded data of this target, such a regular trend can not be due to only noises. We discuss about this target in the next section.

The DWD system J2132$+$0754 has the TESS ID $2000073295$. The maximum power in its LS periodogram occurs at a period of $1.005455$ days which is four times the reported orbital period for this DWD system. Hence, we fold its TESS data into its orbital period $T=0.25056$ days, as shown in the fifth panel of Figure \ref{fig2}. When using the SSA technique to de-noise the data, we re-generate time-series data based on the first $50$ principal components with the lowest correlations with other components. During one orbital period two sinusoidal-like variations occur. Therefore, the period of this trend is half of the orbital period with the difference $\simeq 0.003$ days (considerably less than the peak's FWHM), and this trend likely represents ellipsoidal variation. The FAP value for this target is $8.82\%$ which manifests the trend is relatively certain.

The DWD system J2151$+$1614 also exhibits a sinusoidal-like trend in its TESS light curve. According to the last periodogram in Figure \ref{fig1}, the most-dominant variation in its light curve occurs with a period of $0.196840$ days which is one-third of the reported orbital period for this target. In the periodogram (the last panel of Figure \ref{fig2}), we highlight the orbital period in green. We found that the optimal number of principal components for the SSA technique is between 40-50. Adding more components removes the regular variations and using fewer components reduces the amplitude of this variation. According to its folded light curve (the last panel of Figure \ref{fig2}) a sinusoidal-like variation with a period of $T/3$ presents in the data. For this target, we obtain the FAP value as $36.47\%$ which is too high and its detected trend is not reliable and may be an artifact. According to the last raw of Table \ref{tab1b} the amplitude of this trend is too small (one order of magnitude smaller than the average photometric error) which causes large FAP.

For other DWD systems, either there were no dominant and periodic trends in data, or their most-dominant trends have periods much longer than the orbital periods. Two examples are J0112$+$1835 and J0056$-$0611. In Section \ref{append1} and in Figures \ref{figap1} and \ref{figap2}, we present their periodograms and folded light curves. The light curve of J0112$+$1835 has a periodic trend with a period of $3.3$ days, while its orbital period is $0.15$ days. Its folded light curve is messy. Additionally, the folded light curve of J0056$+$0611 shows a regular periodic trend, but its period is $4.4$ times larger than its orbital period.

\noindent In the next section, we discuss on the origins of the periodic trends shown in Figure \ref{fig2}, which are due to the first six DWD systems in Table \ref{tab1} and put other targets aside.  
\begin{figure*}
    \centering
    \includegraphics[width=0.48\textwidth]{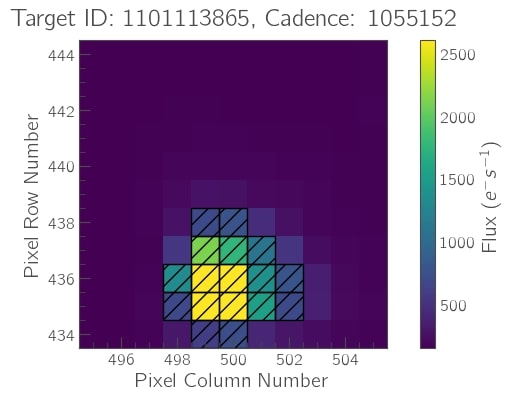} 
    \includegraphics[width=0.48\textwidth]{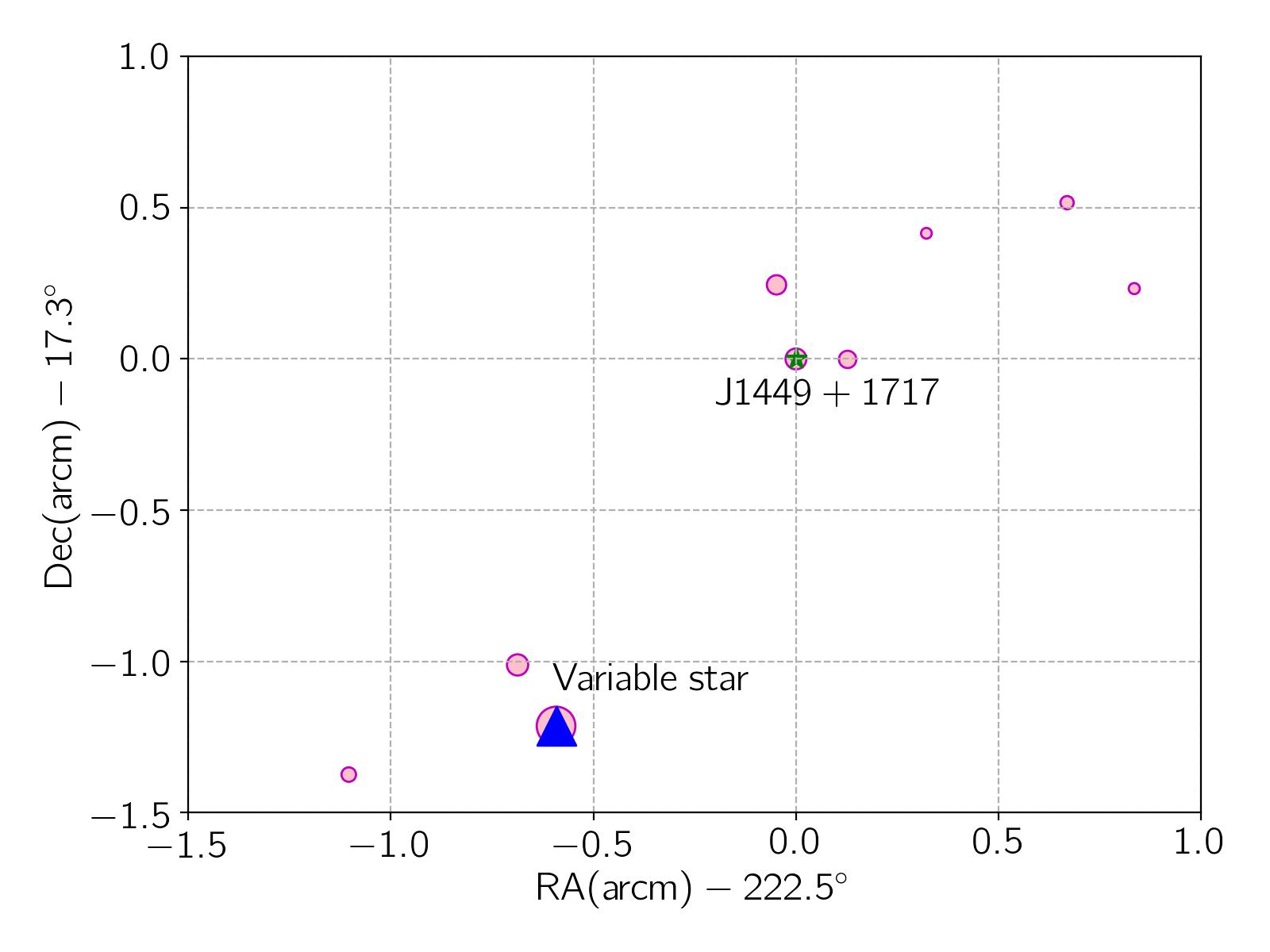}
    \caption{Left panel: The target pixel file drawn from the TESS data at the position TIC 1101113865. The flux of this target is extracted from pixels specified with a dashed pattern. Right panel: The map of blending stars around this target (TIC 1101113865 or J1449$+$1717 as identified with a green star) extracted from the Gaia data archive. The variable and bright star in this field is marked by a blue triangle.}\label{fign}
\end{figure*}  
  
\section{Discussion on detected trends}\label{sec4}
The detected trend from the TESS data for J1449$+$1717 has a large amplitude $\sim 0.7$ mag. This trend was not realized in previous observations. Hence, it is necessary to investigate its origin. The blending effect owing to a close, bright and variable star can make such a trend in the target's light curve, because the TESS pixel size is $21$ arcs and too large. We investigate this hypothesis using the Gaia data archive, and extract a small catalog of the Gaia stars within a circle around this target with the radius $100$ arcs. The target pixel file extracted from the TESS archive reveals that the target light is extracted from an area with the diameter equals to $\sim 5$ pixels as shown in the first panel of Figure \ref{fign}. One of these neighbored stars is variable and very bright. Its Gaia ID is $1236249956118497152$ and it is located at the distance $\sim 80$ arcs from the target. It is brighter than the target by $8.2$ mag in the Gaia $G$-band. We show the map of the Gaia stars around J1449$+$1717 in the second panel of Figure \ref{fign}. In this map, that variable star is specified by a blue triangle. We depicted its periodogram whose maximum power happens at the period $0.36$ days, same as the period of the detected trend for this target. Hence, we put aside this target.

According to the previous section, photometric periodic variations in light curves of J1557$+$2823 and J2151$+$1614 likely have non-orbital origins and could be noises. Because, their FAP values are $14.83$ and $36.47\%$, respectively. These FAP values reveal that their trends are likely artifact and unreliable. Additionally, the periods of their trends are larger than, and one-third of their orbital periods, respectively. So, they could not be made by ellipsoidal variations or Doppler boosting.

\begin{figure*}
\centering
\includegraphics[width=0.48\textwidth]{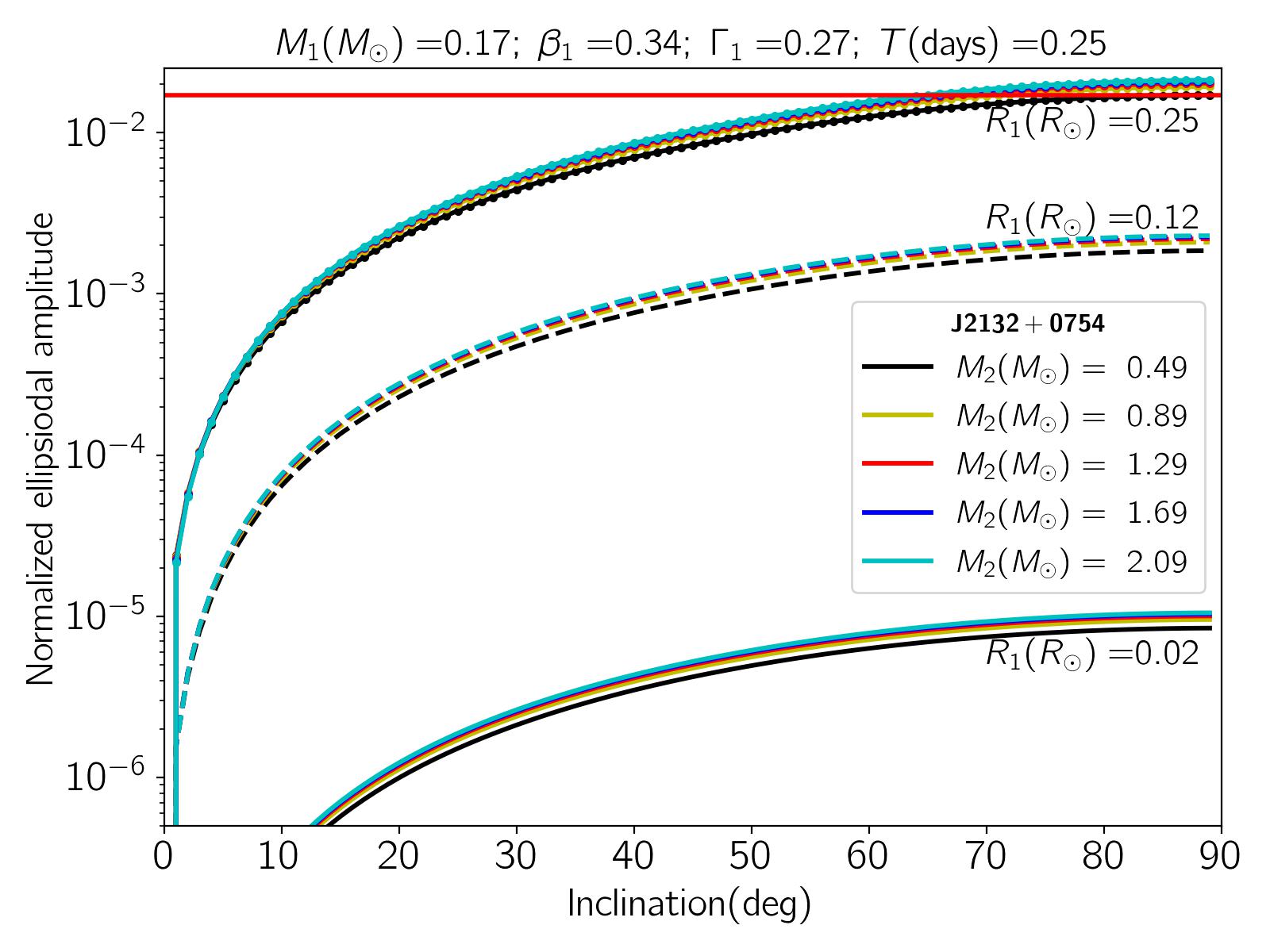}
\includegraphics[width=0.48\textwidth]{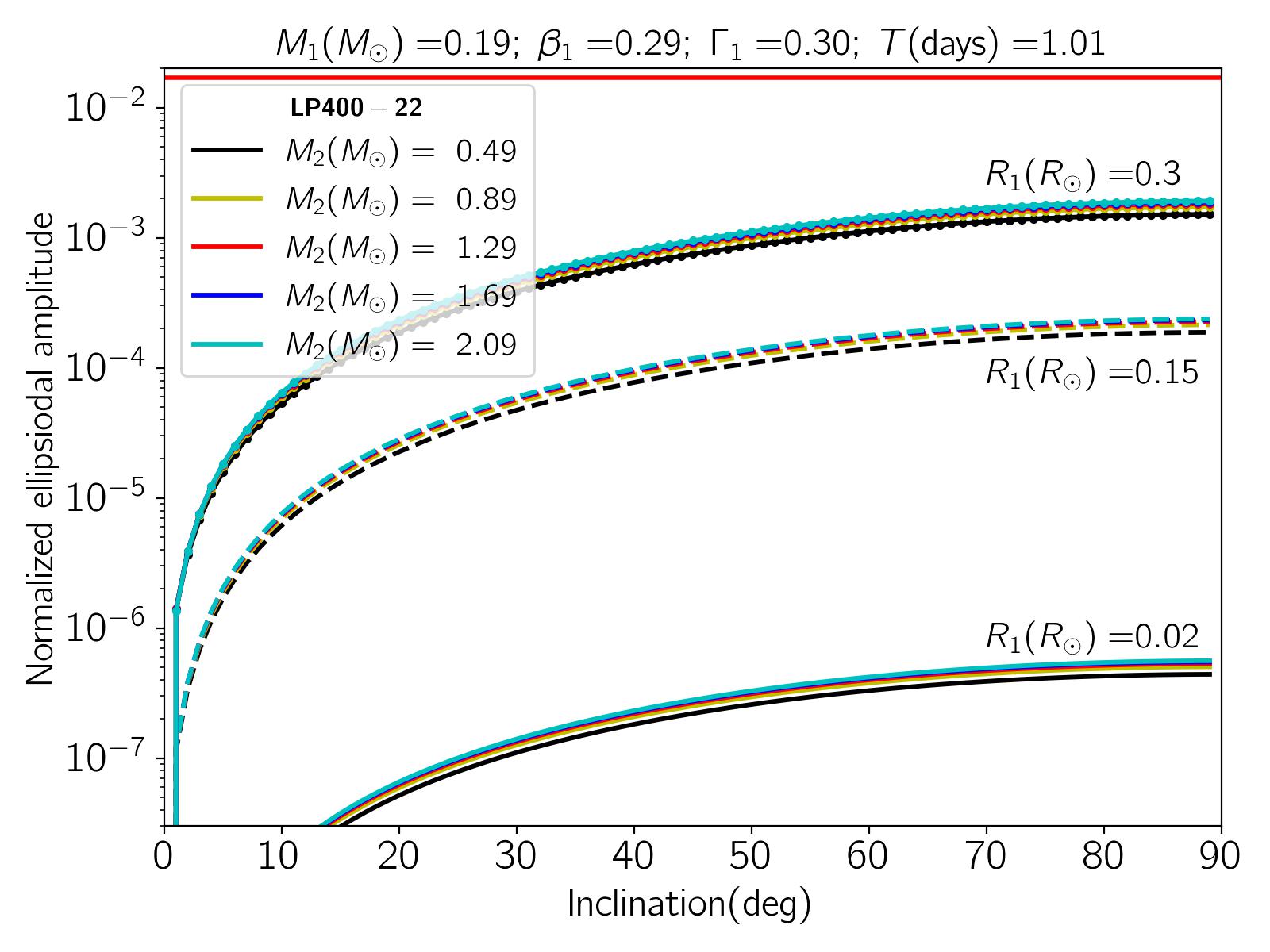}
\caption{The amplitudes of ellipsoidal variations for two targets J2132$+$0754 and LP400$-$22 vary with the inclination angle and by considering different values for the mass of the secondary components (represented by different colors) and the radius of the primary components (represented by different line styles). The thick red lines represent the amplitudes of the observed periodic trends in their TESS light curves $\Delta_{\rm n}$. We utilize the analytical relation provided by \citet{1993ApJEllipsoidal} to evaluate the amplitude of the ellipsoidal variation.}\label{fig3}
\end{figure*}
\begin{deluxetable*}{c c c c c c c c c c c c c}
\tablecolumns{13}
    \centering
    \tablewidth{0.95\textwidth}\tabletypesize\footnotesize
    \tablecaption{The physical parameters of two WD companions in the given DWDs used to simulate their light curves and ellipsoidal variation and Doppler boosting amplitudes.\label{tab2}}
    \tablehead{\colhead{Name}&\colhead{$M_{1}$}&\colhead{$R_{1}$}&\colhead{$T_{\rm{eff},1}$}&\colhead{$\log_{10}[g_{1}]$}&\colhead{$\Gamma_{1}$}&\colhead{$\beta_{1}$}&\colhead{$M_{2}$}&\colhead{$R_{2}$}&\colhead{$T_{\rm{eff},2}$}&\colhead{$\log_{10}[g_{2}]$}&\colhead{$\Gamma_{2}$}&\colhead{$\beta_{2}$}\\ 
        & $(M_{\odot})$ &  $(R_{\odot})$ & $(\rm{K})$ & $(\rm{cm}~s^{-2})$ & & & $(M_{\odot})$ & $(R_{\odot})$ & $(\rm{K})$ & $(\rm{cm}~s^{-2})$& & }
    \startdata    
    J1717+6757&0.185& 0.10 & 14900 & 5.67 & 0.2564 & 0.3379& 0.9 & 0.009 & 15500 & 8.22 & 0.2571 & 0.3168 \\ 
    LP400-22    & 0.19 & -- & 11140 & 6.42 & 0.3045 & 0.291&  -- & 0.014& 10000 &7.8 & 0.3117 & 0.5465\\
    J2132+0754&0.17 & -- & 13700 & 6.0 &  0.2712& 0.3408 & -- &0.004& 15000 & 9.2 & 0.2378 & 0.2235 \\
    \enddata
    \tablecomments{}
\end{deluxetable*}

The light curves of J2132$+$0754 and LP400$-$22 exhibit periodic sinusoidal-like variations with periods equal to and half of orbital periods. Their FAP values are less than 10 per cent (i.e., $8.82\%$ and $3.68\%$, respectively). Hence, their trends are relatively secure. We evaluate the amplitudes of ellipsoidal variations in these systems. \citet{1993ApJEllipsoidal} provided an analytical relation for the ellipsoidal variation, indicating that its amplitude strongly depends on the inclination angle, masses of the two components and the orbital period.

\noindent In Figure \ref{fig3} we display the amplitude of the ellipsoidal variations as a function of the inclination angle, considering different values for the mass of the secondary component (with different colors), and three values for the radius of the primary component (different line styles) in the DWD systems J2132$+$0754 and LP400$-$22. The amplitudes of their detected trends, i.e., $\simeq 0.017$ in normalized flux, are shown with red thick lines in two panels of this figure. The mass of their low-mass companions and their orbital periods are fixed to the values extracted from spectroscopic observations. We estimate the limb-darkening $\Gamma$ and gravity-darkening $\beta$ coefficients of two components based on their surface temperatures and surface gravities \citep{2020AAClaret}. All applied parameters to generate ellipsoidal amplitudes for these targets are mentioned in Table \ref{tab2}.

\begin{figure}
\centering
\includegraphics[width=0.48\textwidth]{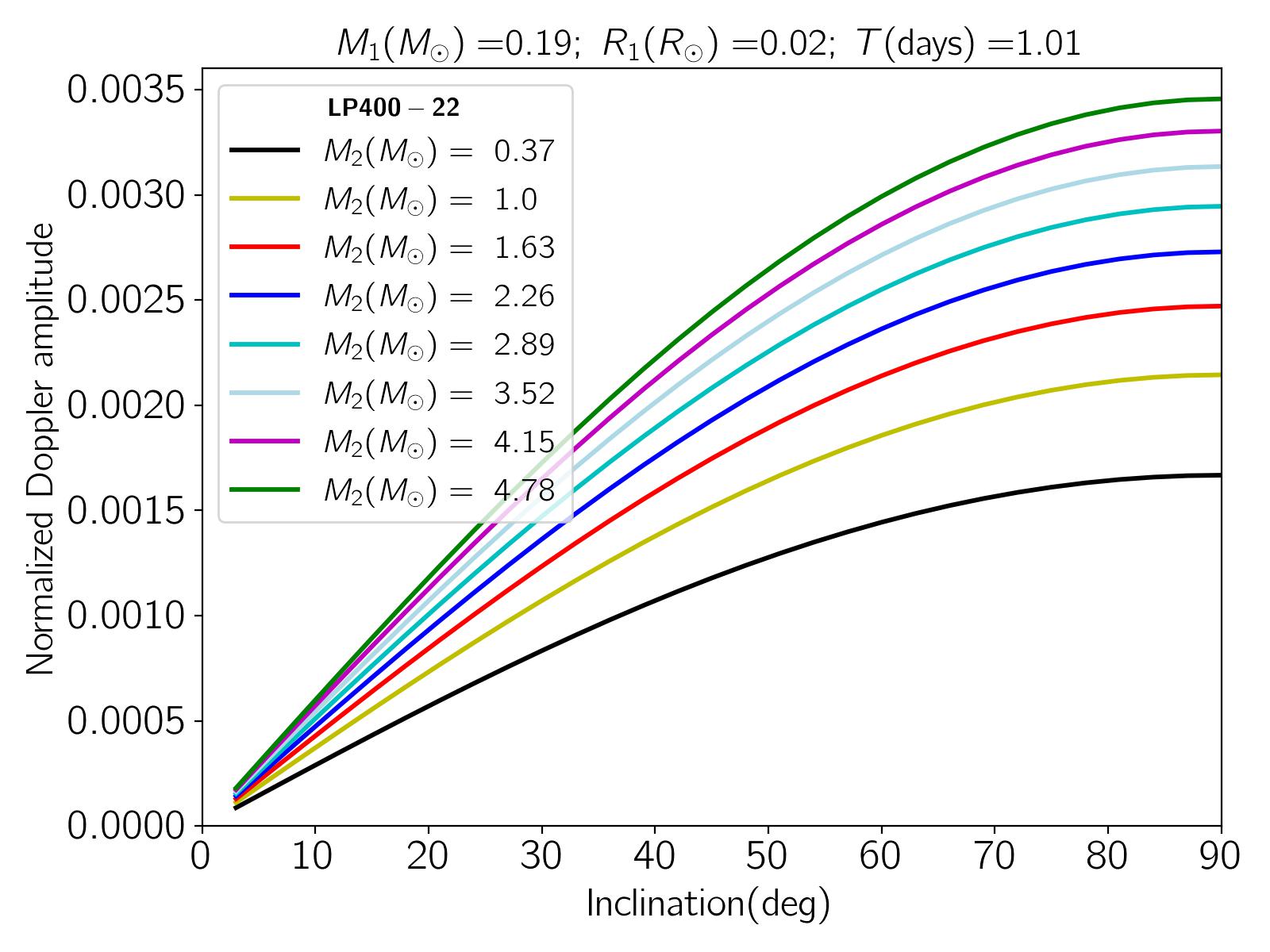}
\caption{The expected amplitude in the normalized flux due to Doppler boosting (given by Equation \ref{eqq}) varies with the inclination angle and when considering different values for the mass of the secondary companion in LP400$-$22. We assume a circular orbit in the simulation.}\label{fig4}
\end{figure}
\begin{figure}
\centering
\includegraphics[width=0.48\textwidth]{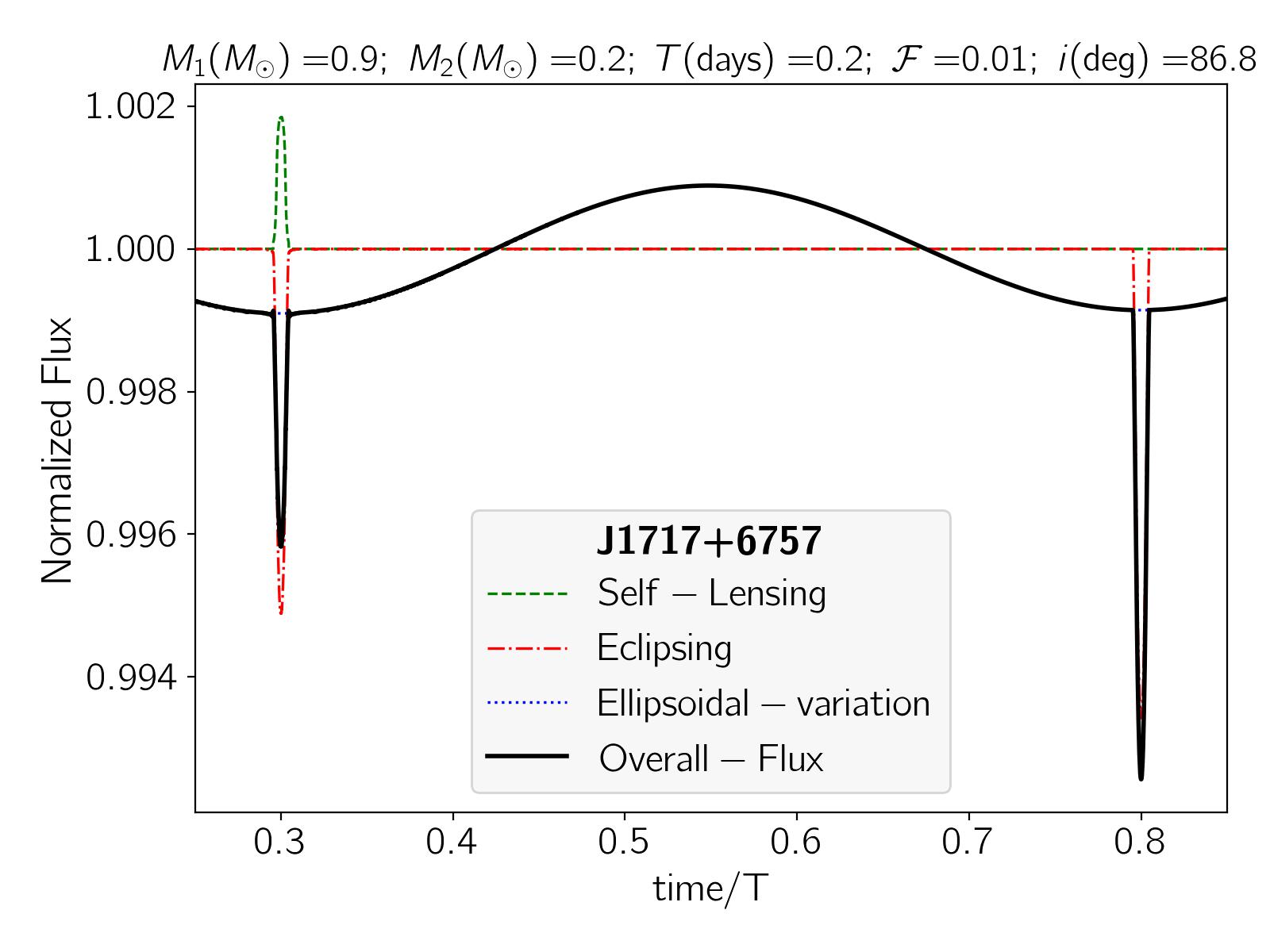}
\caption{The expected light curve for the DWD J1717$+$6757 taking into account self-lensing (plotted by dashed green curve), eclipsing effect (dot-dashed red curve), and ellipsoidal variation (dotted blue curve). The overall flux due to the two components is shown by the solid black curve. The parameters used to create this light curve are reported in the first row of Table \ref{tab2}.}\label{fig5}
\end{figure}


According to the first panel of Figure \ref{fig3}, the maximum possible amplitudes for ellipsoidal variations for the first target is $\sim 0.02$ (when $R_{\rm 1}\simeq 0.25 R_{\odot}$). Therefore, the periodic trend in the J2132$+$0754 light curve will be the ellipsoidal variation if its primary component has a radius $\simeq 0.25 R_{\odot}$ which is larger than that for a typical WD with a larger mass. We note that \citet{Hermes_2014_ellipsiodal} have done dense photometric observations from several DWD systems including ELM WDs in close orbits and captured ellipsoidal variations for eight ones. They combined the results from photometric and spectroscopic observations from these systems and determined the radii of ELM WDs. Their Figure (5) shows that ELM WDs' radii reach to $0.18 R_{\odot}$. Hence, ELM WDs in close orbits with massive and compact objects are bigger than isolated ones which somewhat justifies our results. Future high-speed photometric observations from this target will help to determine the nature of its trend.
    
\noindent Besides, the second panel of Figure \ref{fig3} manifests the ellipsoidal variation does not create the identified sinusoidal-like trend in photometric data of LP400$-$22, because its possible amplitudes are significantly smaller than that detected.
    
\noindent The sinusoidal trend detected in the TESS light curve LP400$-$22 has the period equals to its orbital period and so it can be generated by Doppler boosting. This effect occurs for binary systems with a close orbit and a massive companion \citep[see, e.g., ][]{2010ApJShporer}. We numerically calculate this effect for LP400$-$22 by considering different inclination angles and the mass of the secondary companion using the following relation:  
\begin{eqnarray}
\Delta(v_{\rm{los}}, T_{\star})= \frac{\int_{0}^{\infty} d\lambda \Big[\mathcal{F}(\lambda', T_{\star})-\mathcal{F}(\lambda, T_{\star}) \Big] \mathcal{T}_{\rm{TESS}}(\lambda) }{\int_{0}^{\infty}d\lambda \mathcal{F}(\lambda, T_{\star}) \mathcal{T}_{\rm{TESS}}(\lambda)}, 
\label{eqq}
\end{eqnarray} 
where, $\mathcal{T}_{\rm{TESS}}(\lambda)$ is the throughput function of the TESS $T$-band filter versus the wavelength $\lambda$, $\mathcal{F}(\lambda, T_{\star})$ is the Planck spectral density, which is a function of the wavelength and the surface temperature of the bright companion. Additionally $\lambda'= \lambda\big(1+v_{\rm{los}}/c\big)$ represents the boosted wavelength due to the motion of the primary companion with the line of sight velocity $v_{\rm{los}}$, and $c$ is the speed of light. We note that $v_{\rm{los}}$ depends on the inclination angle of the orbital plane $i$. The amplitude of this Doppler boosting for the target LP400-22 is shown versus the inclination angle and for different values for the mass of the secondary companion in Figure \ref{fig4}. Accordingly, Equation \ref{eqq} predicts the amplitude of the Doppler boosting to reach $0.004$ in the normalized flux, while the amplitude of the detected trend in the light curve of LP400-22 is $\simeq0.017$. For calculating the amplitude of the Doppler boosting we have made some simplifying assumptions which could make this underestimation. These assumptions are listed here: (i) We ideally assume the Planck distribution as the spectral density of a ELM WD $\mathcal{F}(\lambda, T_{\star})$, while its true spectral density is extracted from spectroscopic observations covering a wide wavelength range. (ii) We set $11,140$ K as the effective temperature of the ELM WD in this system. (iii) We ignore the flux due to the faint companion in Equation \ref{eqq}. (iv) We assume a circular orbit for the ELM WD around its massive companion. (v) Additionally, the observed trend in the light curve of LP400$-$22 could be disturbed by noises because its FAP is $3.68\%$. (vi) There are some other possible sources of variations that could change the Doppler boosting's amplitude such as stellar spots \citep[e.g., see, ][]{2022ApJSKnote,2015MNRASBalaji}, time-varying contamination by reflecting lights, existence of hot spots due to mass transfer. To resolve this problem, one needs spectral densities of two companions through spectroscopic observations in a wide range of wavelength.

For the target J1717$+$6757 we simulate its light curve by considering self-lensing, eclipsing effects and ellipsoidal variations as shown in Figure \ref{fig4}. In this plot, these effects over time are represented by dashed green, dot-dashed red, and dotted blue curves. The overall flux is shown by solid black curve. In this plot, during its first signal, the massive WD with a mass of $0.9 M_{\odot}$ is the lens which creates a considerable lensing effect. In the second signal, the massive WD is the source star, and a deep eclipse happens. We note that the massive WD flux is only $0.01$ of the less massive one. As a result, the durations of the two lensing/eclipsing signals are as short as $3$ minutes, while the TESS cadence for observing this target was 10 and 30 minutes. Hence, this telescope was unable to capture the eclipsing/lensing signals for this target.

\section{Conclusions}\label{sec5}
Recently, several spectroscopic surveys (such as ELM, Gaia, and LAMOST) have identified a significant number of DWD systems  through precise radial velocity measurements of bright companions, which most of them contain at least one ELM WD. Our universe is not old enough for ELM WDs to form through the evolution of single low-mass stars. Instead, they are found in close orbits with massive companions. DWD systems are considered potential progenitors of Type Ia supernovae and are also believed to be sources of gravitational waves. Given the importance of these systems, we searched photometric TESS data to find any regular and periodic signals/trends in their light curves. Our initial dataset included 67 known DWD systems, and TESS data were available for 17 of them which were listed in Table \ref{tab1}.  

For these seventeen DWD systems, we extracted their light curves and de-noised them using the SSA technique. We searched for their most-dominant principal and periodic components by plotting periodograms. For six systems we could find some periodic trends in their light curves. These systems were J1717$+$6757, J1557$+$2823, LP400$-$22, J1449$+$1717, J2132$+$0754, and J2151$+$1614. 

The periodic trends in the TESS light curves of J1449$+$1717, J1557$+$2823, and J2151$+$1614 had non-orbital origins. The periodic trend in the first target had unexpectedly a large amplitude with an ignorable FAP value. Thanks to the Gaia catalog around this target, we found that the nature of this variation is a variable and very bright star at the distance $80$ arcs from the target which were unresolvable in the TESS target pixel file (see Figure \ref{fign}). The TESS light curve of J1557$+$2823 shows a periodic trend with the period longer than the orbital period. Its FAP value was $14.83\%$.  Hence, its periodic trend is not secure and could be an artifact. There was a periodic trend in the light curve of J2151$+$1614 with the period equals to one-third of its orbital period, but its FAP was $36.47\%$. This high value of FAP means that this trend is likely originated from noises and unreliable.

\noindent For the target J1717$+$6757, the TESS data well recovered its periodic trend (with the period exactly equal to the orbital period) as previously was detected by \citet{Vennes2011J17176757}. For this target \citet{2014MNRASHermesJ17176757} detected two eclipsing signals. These signals were not detectable in its light curve, because of the sparse TESS data taken for this event.

For the targets  J2132$+$0754 and LP400$-$22, we found sinusoidal-like trends in their light curves with periods half of and equals to their orbital periods, respectively. These variations are producible by ellipsoidal variation and Doppler boosting. The first effect is owing to the re-shaping of non-compact or low-mass objects by the strong gravitational forces of their compact and massive companions. The second is caused by the periodic motion of a low-mass companion in the line of sight direction, with a period equal to the orbital period. We calculated FAP values for their light curves as $8.82\%$ and $3.68\%$, which reveal that these trends are relatively certain. We estimated the amplitudes of ellipsoidal variations in these systems by considering different amounts for mass of their massive and faint companions, the inclination angle, and the radius of their primary companions based on the theoretical relation provided by \citet{1993ApJEllipsoidal}. For J2132$+$0754, the observed sinusoidal-like trend is the ellipsoidal variation if the primary companion has a large radius $\simeq 0.25 R_{\odot}$. 

We evaluated the amplitude of Doppler boosting for the target LP400$-$22 by exploring various values for the inclination angle and the mass of the secondary companion. The observed periodic trend in its TESS light curve is higher than the simulated ones based on Equation \ref{eqq}. For precise modeling this trend, one needs spectral densities and effective temperatures of two companions for this target.

Our study highlights the ability of TESS observations in identifying the periodic trends in the light curves of DWD systems with periods ranging from several hours to a few days. These periodic signals could be generated by ellipsoidal variations, Doppler boosting, intrinsic variations, or flares. Nevertheless, the large size of the TESS pixel enhances the blending effect especially for faint objects, e.g., J1449$+$1717. We could not discern eclipsing or lensing signals in the photometric light curves of known DWDs. This is primarily due to the TESS cadences for these faint objects (all of them were in the TESS FFIs) which are 3.3, 10 and 30 minutes, while the duration of lensing/eclipsing signals in DWDs is approximately one minute. Therefore, TESS data is very valuable for finding main trends in the data of DWD systems with periods spanning several hours to a few days specially for bright targets with ignorable blending effects.
  	
\small{The source codes have been developed for this work can be found in the GitHub and Zenodo addresses: \url{https://github.com/SSajadian54/RealDWDSystem}, and \url{https://zenodo.org/records/15867446} \citep{sajadian2025Zenodo}. }\\

\small{This paper includes data collected by the TESS mission. We used the light curves of seventeen FFIs' stars extracted by the \texttt{TESS}-\texttt{SPOC} pipeline (with DOI number: \url{doi:10.17909/t9-wpz1-8s54}) which were collected by the TESS mission that are publicly available from the MAST. Funding for the TESS mission is provided by NASA's Science Mission directorate. We acknowledge the use of TESS Alert data, which is currently in a beta test phase, from pipelines at the TESS Science Office and at the TESS Science Processing Operations Center. This work has also made use of data from the European Space Agency (ESA) mission Gaia (\url{https://www.cosmos.esa.int/gaia}), processed by the Gaia Data Processing and Analysis Consortium (DPAC, \url{https://www.cosmos.esa.int/web/gaia/dpac/consortium}). Funding for the DPAC has been provided by national institutions, in particular the institutions participating in the Gaia Multilateral Agreement.}
\small{The authors gratefully thank the anonymous referee for his/her careful and helpful comments and suggestions.}

\appendix
\section{The TESS light curve and periodogram of  J0112$+$1835 and J0056$-$0611}\label{append1}
\begin{figure*}
\centering
\includegraphics[width=0.48\textwidth,height=0.2\textheight]{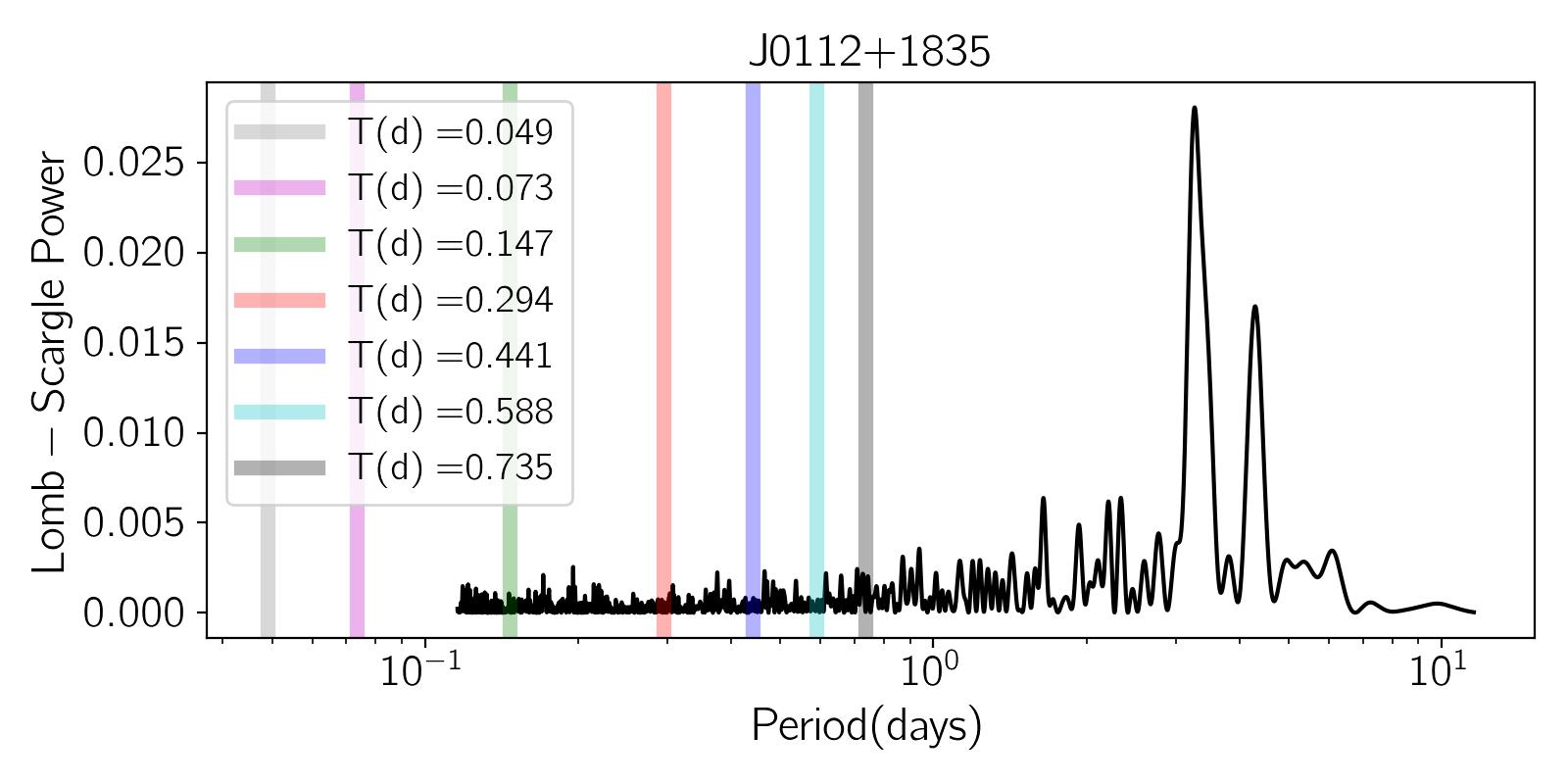}
\includegraphics[width=0.48\textwidth,height=0.2\textheight]{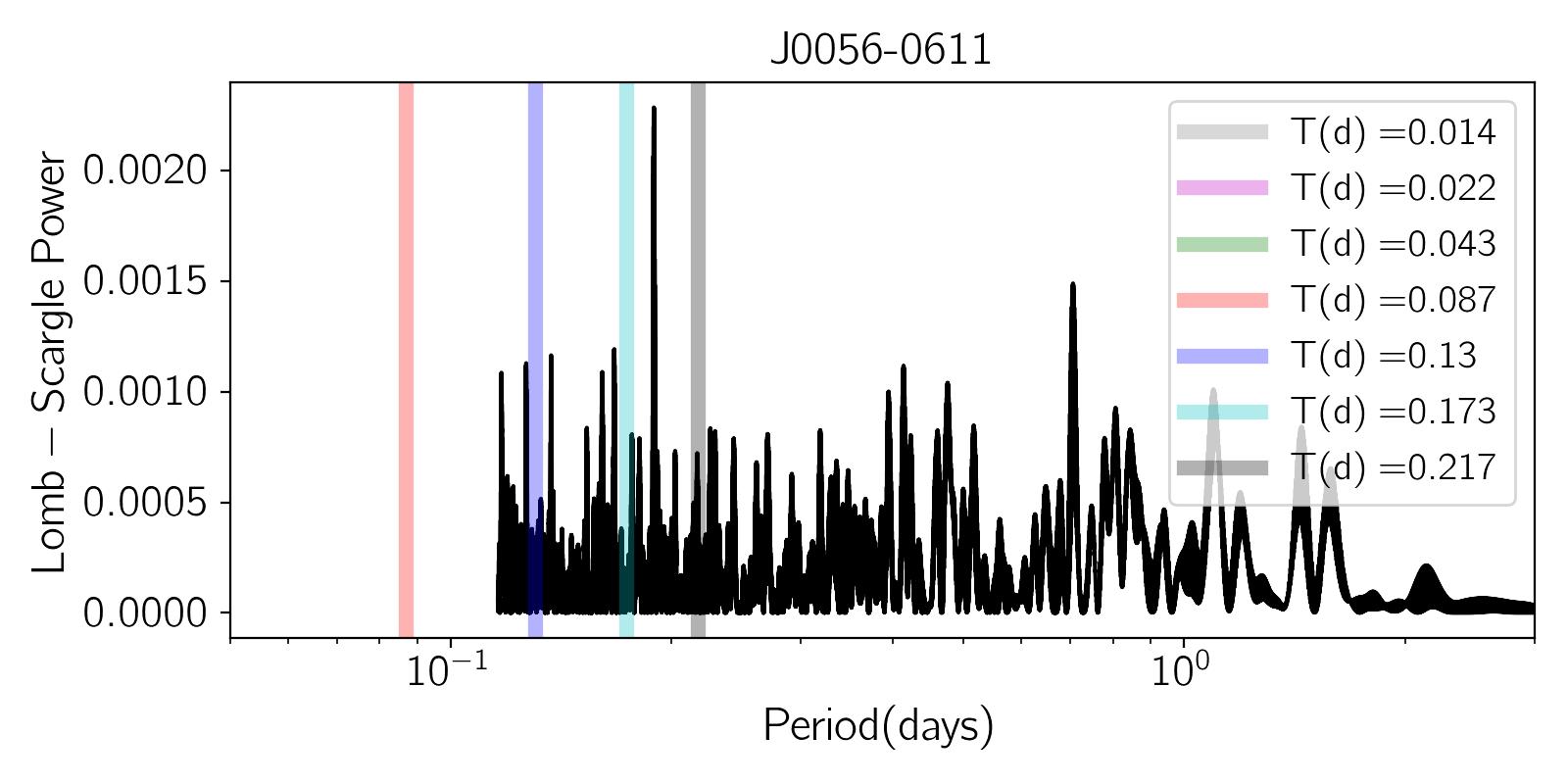}
\caption{Two panels show the LS periodograms related to the TESS light curves of DWD systems J0112$+$1835 (left panel) and J0056$-$0611 (right panel). The colored and thick lines are plotted at their orbital periods, and proper fractions or multiples of their orbital periods.}\label{figap1}
\end{figure*}

\begin{figure*}
\centering
\includegraphics[width=0.48\textwidth]{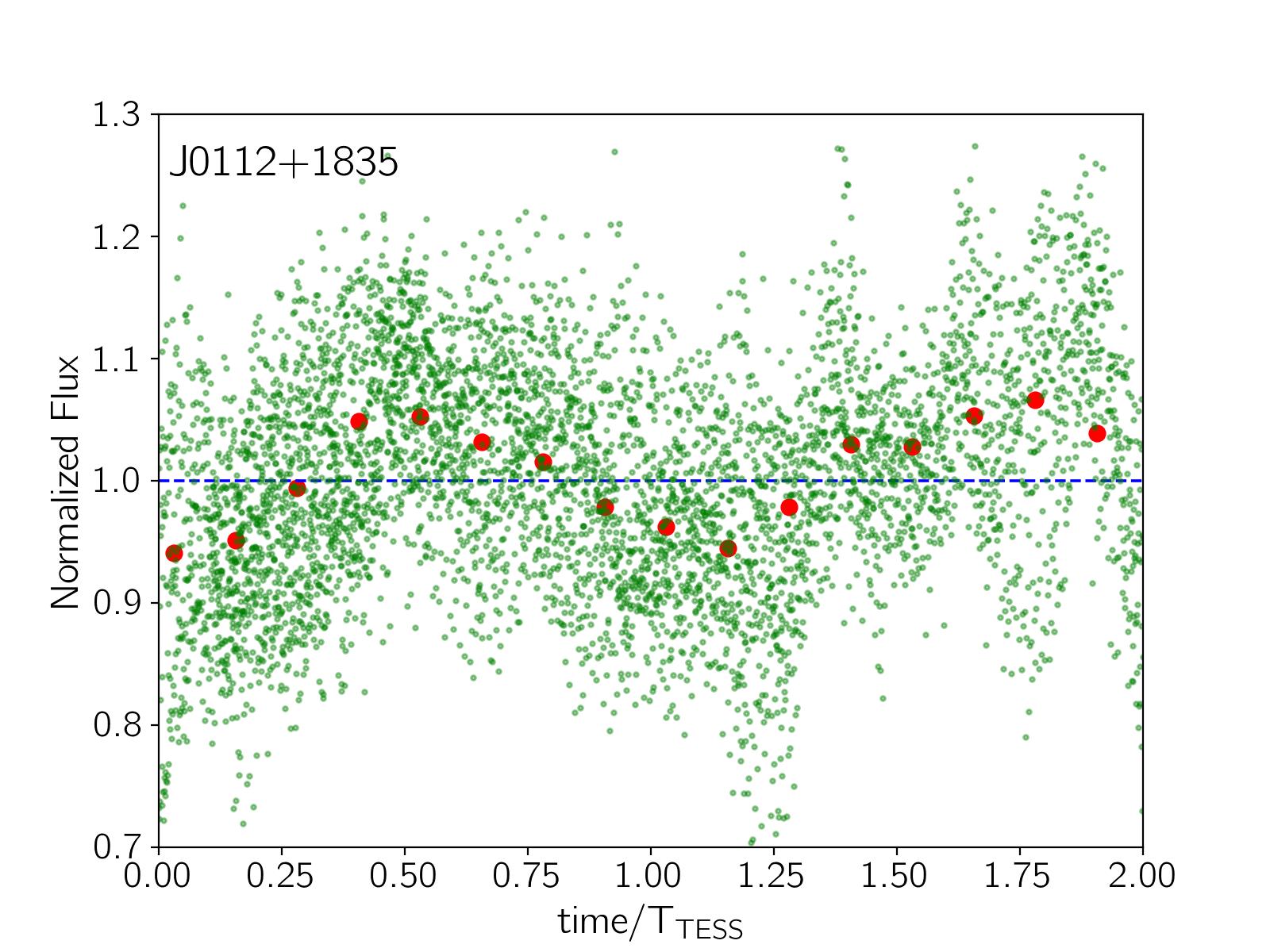}
\includegraphics[width=0.48\textwidth]{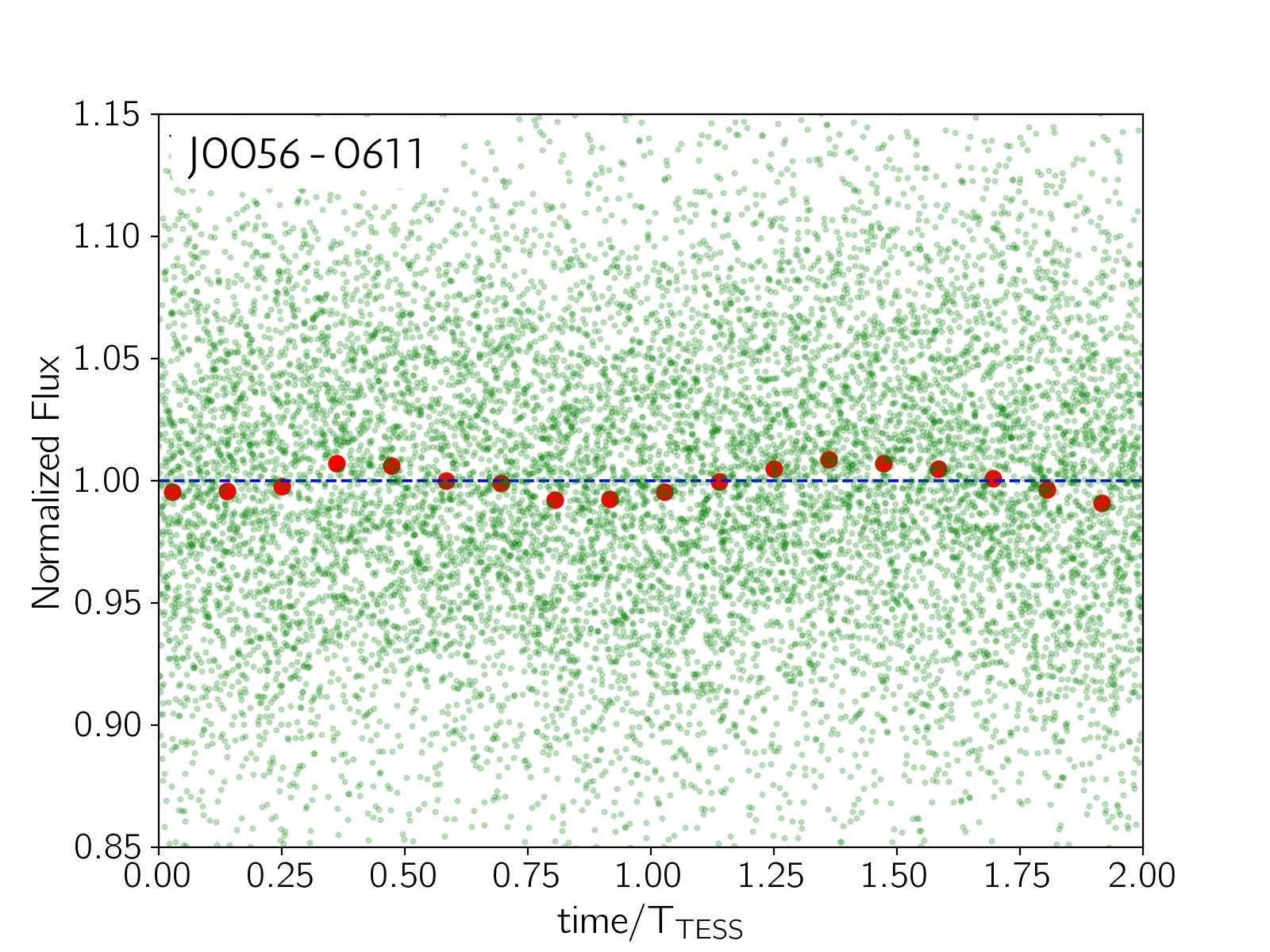}
\caption{The folded and de-noised light curves due to two DWD systems J0112$+$1835 (left panel) and J0056$-$0611 (right panel), respectively. $T_{\rm{TESS}}$ values are periods of trends with the maximum LS powers (mentioned in the sixth column of Table \ref{tab1}).}\label{figap2}
\end{figure*}

The periodograms and folded light curves for two DWD systems,  J0112$+$1835 and J0056$-$0611 with the TESS IDs 611358291 and 610678231, are displayed in Figures \ref{figap1} and \ref{figap2}, respectively. Their $T_{\rm{TESS}}$ values representing the periods at which peaks of their LS powers occur are much longer than their orbital periods.  

The folded light curve due to J0112$+$1835 shows drastic variations like flares. But the periodic trend in the light curve J0056$-$0611 is regular and has sinusoidal-like shape.  
\bibliography{ref}{}
\bibliographystyle{aasjournal}
\end{document}